\documentclass[11pt]{article}
\usepackage{jheppub}
\usepackage{epstopdf}
\usepackage{enumitem}
\usepackage{tcolorbox}
\usepackage{cite}
\usepackage{jheppub}
\usepackage{mathrsfs}
\usepackage{psfrag}
\usepackage{color}
\usepackage{hyperref}
\allowdisplaybreaks

\usepackage{slashed}
\usepackage{feynmp-auto}
\usepackage{simplewick}
\usepackage{cancel}

\usepackage{todonotes}

\usepackage{amsmath}
\usepackage{amsfonts}
\usepackage{graphicx}
\usepackage{amssymb}
\usepackage{xcolor}

\usepackage{mathtools}

\usepackage{bm}

\usepackage{tikz-feynman}

\DeclarePairedDelimiter\ceil{\lceil}{\rceil}
\DeclarePairedDelimiter\floor{\lfloor}{\rfloor}

\usepackage{amsmath,bbm,array,amsfonts,graphicx,wrapfig,arydshln,lscape,float,multirow,longtable,rotating,makecell}
\usepackage{url}

\DeclareMathAlphabet\mathbfcal{OMS}{cmsy}{b}{n}

\newcommand{\eq}{\begin{equation}}
\newcommand{\eqe}{\end{equation}}
\newcommand{\eqa}{\begin{eqnarray}}
\newcommand{\eqae}{\end{eqnarray}}

\newcommand{\bn}{\begin{enumerate}}
\newcommand{\en}{\end{enumerate}}

\def\r{\rho}

\newcommand{\bfig}{\begin{figure}}
\newcommand{\efig}{\end{figure}}

\def\abs#1{{\left| #1 \right|}}

\def\ie{\begin{equation}\begin{aligned}}
\def\fe{\end{aligned}\end{equation}}

\def\bl#1\el{\begin{align} #1 \end{align}}
\def\bg#1\eg{\begin{gather} #1 \end{gather}}

\def\bld#1\eld{\begin{aligned} #1 \end{aligned}}
\def\bgd#1\egd{\begin{gathered} #1 \end{gathered}}

\newcommand{\ket}[1]{|{#1}\rangle}

\renewcommand{\tt}{\texttt}

\newcommand{\AB}[1]{\langle #1 \rangle}
\newcommand{\SB}[1]{[ #1 ]}

\newcommand{\BS}[1]{\boldsymbol{#1}}
\newcommand{\RAB}[1]{| #1 \rangle}

\newcommand{\RSB}[1]{| #1 ]}

\newcommand{\RN}[1]{%
  \textup{\uppercase\expandafter{\romannumeral#1}}%
}

\newcommand*{\mathcolor}{}
\def\mathcolor#1#{\mathcoloraux{#1}}
\newcommand*{\mathcoloraux}[3]{%
  \protect\leavevmode
  \begingroup
    \color#1{#2}#3%
  \endgroup
}

\title{Minimal spin deflection of Kerr-Newman and Supersymmetric black hole}
\author{Bo-Ting Chen$^{1}$}
\author{Ming-Zhi Chung$^{1}$}
\author{Yu-tin Huang$^{1,2}$}
\author{Man Kuan Tam$^{1}$}

\affiliation{$^1$ Department of Physics and Center for Theoretical Physics, National Taiwan University, Taipei 10617, Taiwan}
\affiliation{$^2$ Physics Division, National Center for Theoretical Sciences, Taipei 10617, Taiwan}

\emailAdd{r08222086@ntu.edu.tw}
\emailAdd{dchung0741@gmail.com}
\emailAdd{yutinyt@gmail.com}
\emailAdd{mankuantam@gmail.com}

\abstract{
Recent studies have shown that minimal couplings for massive spinning particles, which in the classical limit reproduce the leading PM Kerr black hole dynamics, leads to an Eikonal S-matrix exhibiting spin-entanglement suppression. In this paper we trace this phenomenon to the suppression of spin-flipping components in the S-matrix, known to be the hallmark of minimal coupling in the ultra-relativistic limit. We further generalize the consideration to charged and $\mathcal{N}=4$ blackholes, demonstrating that in both cases maximal suppression occurs at the extremal limit. }

\begin{document}
\maketitle

\section{ Introduction }
It has long been known that the dynamics in general relativity can be extracted from the scattering of particles. This was first realized in the context of  computations of the conservative gravitational potential~\citep{Iwasaki:1971vb, Iwasaki:1971iy, Donoghue:1994dn, BjerrumBohr:2002kt, Bjerrum-Bohr:2013bxa, Neill:2013wsa, Cristofoli:2019neg}, and in recent years extended to linear/angular impulse~\citep{Kosower:2018adc, Maybee:2019jus}, scattering angle~\citep{Kalin:2019rwq, Bjerrum-Bohr:2019kec, Kalin:2019inp}, radiation effects~\citep{DiVecchia:2020ymx, DiVecchia:2021ndb, Herrmann:2021tct}  as well as spin effects-\citep{Guevara:2017csg, Guevara:2018wpp, Bern:2020buy, Kosmopoulos:2021zoq}.  The combination of modern approaches to scattering amplitude~\citep{Cheung:2018wkq} and the effective field theory description of gravitational wave observables~\citep{Goldberger:2004jt} (see~\citep{Bern:2019crd} for summary), has culminated into the state of the art third~\citep{Bern:2019nnu}  and fourth post-Minkowskian order potential~\citep{Bern:2021dqo}.

The remarkable effectiveness of the particle description for black hole dynamics invites us to ask: when viewed through the prism of scattering amplitudes do we find hidden features of black holes? An intriguing aspect of this question is that we asking if there are special features of black hole solutions that are visible at long distances, $r\gg r_s$. As Birkhoff's theorem reduces any spherically symmetric solution to Schwarzschild, this question becomes nontrivial when we have spins, i.e. the Kerr solution. Indeed even at leading order in $G_N$, the post Minkowskian expansion (PM), rotating bodies induces an infinite number of gravitational multipole moments \citep{Porto:2008jj, Levi:2015msa}, for which the Kerr solution dictates a unique value \citep{Vines:2017hyw}. That these values are special were supported by its matching to the minimal coupling of spinning particles
\citep{Guevara:2017csg, Chung:2018kqs, Guevara:2018wpp, Arkani-Hamed:2019ymq, Aoude:2020onz}, which are defined kinematically for the matrix elements between states with fixed mass and spins. In fact, the infinite number of multipoles were reproduced by a single coupling in the kinematic representation.

Motivated by the observation that symmetry enhancement in nucleon scattering are often accompanied by spin-entanglement suppression, the authors analyzed the spin entanglement entropy of $2\rightarrow 2$ $S$-matrix for minimal coupling in \citep{Aoude:2020mlg}. More precisely, starting with a given in-state in the spin Hilbert space, which can be either pure or mixed, one asks the change in spin entanglement entropy of the outstate. Naively, due to the interaction in the $S$-matrix, one would expect that the out-state incurs extra entanglement. In the case of strong interaction, it was observed that the entanglement was suppressed for the spin-flavor symmetry enhancement point \citep{Beane:2018oxh}. By considering the Eikonal $S$-matrix  parameterized the worldline Wilson coefficients (encoding the gravitational multipole moments) of one-particle EFT~\citep{Porto:2005ac, Porto:2008jj}, it was found that the ``relative spin-entanglement" between in- and out-state vanishes for minimal coupling, i.e. with the Wilson coefficients set to Kerr values. Thus the interaction whose classical limit reproduces the dynamics of rotating black-holes at leading order in PM, generates near zero entanglement for spinning particles.

In this paper, the goal is two fold: 1. to seek the origin of spin-entanglement suppression for minimal coupling, and 2. to observe such behaviour for rotating black hole solutions with tunable parameters, such as charge (central charge). First, by taking the ultra-relativistic limit of the Kerr Eikonal $S$-matrix, we observe that the Eikonal phase in spin space takes the form 
\begin{equation}
\left.\chi\right|_{k\gg 1} = \small
k^2\left(
\begin{array}{cccc}
 4 \log b & 0 & \frac{4 \left(C_{2}-1\right)}{b^2m^2} & \cdots 
   \\
 0 & 4  \log b & 0 & \cdots  \\
 \frac{4 \left(C_{2}-1\right)}{b^2m^2} & 0 & 4  \log b & \cdots   \\
 \vdots &  \vdots &  \vdots &  \vdots\\
 \end{array}
\right) +\mathcal{O}(k)
\end{equation}
where $k=\frac{|\vec{p}|}{m}$, the ratio of the center of mass (spatial)momenta with respect to mass, $b$ the impact parameter and $C_{2}$ the Wilson coefficients for which Kerr solution corresponds to $C_n=1$. As the off-diagonal terms corresponds to the change in spin state, we see that in the high energy limit, the Kerr solution leads to vanishing spin flipping components. This naturally explains the observed suppression in the relative spin-entanglement entropy. Note that minimal coupling is uniquely determined kinematically by requiring that in the high energy limit (massless limit), only helicity preserving interactions remain.

We further generalize to cases where the black hole solutions admit a tunable parameter: the Kerr-Newman and the non-BPS solution in $N{=}4$ supergravity. For the former, the tunable parameter is the charge while for the latter is the central charge. For Kerr-Newman, it was shown in \citep{Arkani-Hamed:2019ymq, Chung:2019yfs}, that the 1 PM spin-multipoles are given by gravitational and electro-magentic minimal coupling. We analyze the corresponding Eikonal phase in the near extremal limit. We find that once again, the spin-flip components are minimized at the extremal limit. 

For $N{=}4$ supergravity, we utilize the non-BPS on-shell scattering amplitude derived in \citep{Chen:2021hjl}, and show by introducing suitable spin-factors reproduces the impulse of a test particle in the non-BPS rotating black hole background. This matching is non-trivial given that the impulse receives contributions from the metric, gauge potential and scalar profile. Equipped with the on-shell amplitude, we compute the 1 PM Eikonal phase. Not surprisingly, we see that the extremal limit corresponds to spin-flip suppression limit.

This paper is organized as follows: In section \ref{sec: General_Eikonal}, we briefly review the entanglement caused by a general scattering process. This requires us to construct the most general form of the eikonal phase, which is the tree level amplitude of two massive particles of spin $s$ exchanging a graviton, photons and complex scalars. Then we restrict ourselves to the pure graviton exchange case and give an explicit explanation of why the gravitational minimal coupling amplitude gives minimal entropy increment. In section \ref{sec: KN}, we turn on the photon minimal coupling amplitude along with tunable charge to mass ratio of the scattering particles. We first show that our amplitude can reproduce the impulse computed from integrating the classical equations of motion. Then we turn to the eikonal phase and then show that the entanglement increment is minimized when extremal limit is satisfied. Finally in section \ref{sec: SUSY}, we construct the 4pt amplitude with graviton, graviphoton and scalar exchange from the $\mathcal{N}=4$ superamplitude with central charge extension derived in \citep{Chen:2021hjl}, where the amplitude is now a function of the central charge. We first match the impulse computed from this amplitude and compare it with that from the classical equations of motion. Then we show that the entropy increment is minimized when the BPS limit is saturated.

\section{ The Eikonal $S$-Matrix and spin entanglement}\label{sec: General_Eikonal}
We consider the entanglement in spin space caused by a long range scattering process. With a given in-state, we can obtain the out state with the eikonal $\mathcal{S}$-matrix:
\begin{equation}
\ket{\text{out}} = \mathcal{S}_{Eik}  \ket{\text{in}} = e^{i\chi} \ket{\text{in}},
\end{equation}
where $\ket{\text{in}}$ and $\ket{\text{out}}$ live in the spin Hilbert space $\mathcal{H} = \mathcal{H}_{s_a} \otimes \mathcal{H}_{s_b}$ . The eikonal phase $\chi$ in the impact parameter space ($\vec{b}$ space) is defined as
\begin{equation}
\chi = \frac{1}{4|\vec{p}|E}\int \frac{d^2 \vec{q}}{(2\pi)^2} e^{i\vec{q}\cdot \vec{b}}\; U_a \otimes U_b M,
\end{equation}
where $M$ is the tree level amplitude and $U_a$, $U_b$ are the Hilbert space matching factors introduced in \citep{Chung:2019duq, Chung:2020rrz} to express the out-state with the basis of the in-state spin Hilbert space. 
\begin{figure}[t]
\centering
\includegraphics[scale=0.4]{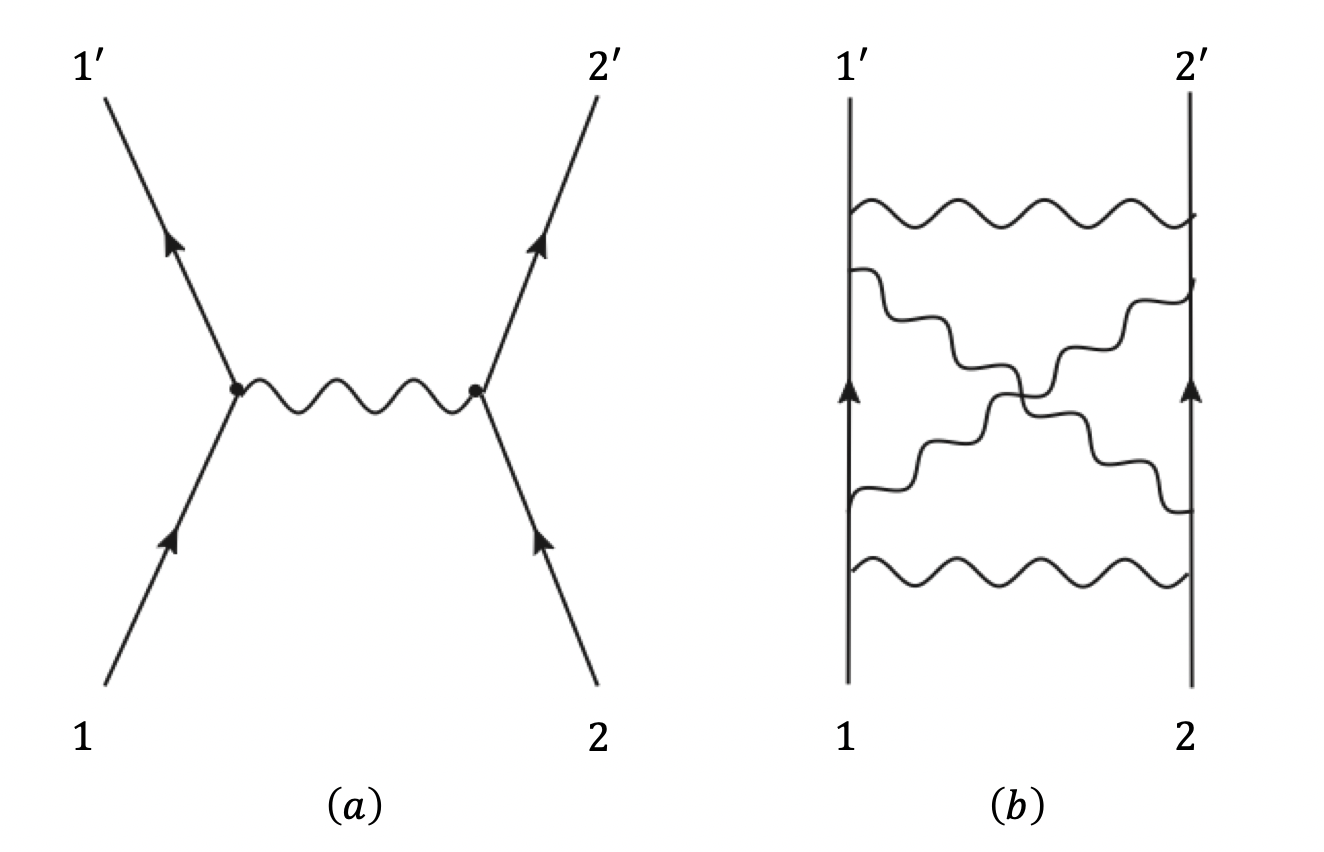}
\caption{
(a) The tree level amplitude which contributes to leading order in Newton constant $G$. (b) The eikonal amplitude which resums all ladder diagrams.
}\label{fig:Scattering_Graph}
\end{figure}
With the eikonal amplitude at hand, we can discuss the difference in the von Neumann entropy between the in-state and out-state
\begin{equation}
\Delta S \equiv - \text{tr}[\rho_{\text{out}} \log \rho_{\text{out}}] +  \text{tr}[\rho_{\text{in}} \log \rho_{\text{in}}],
\end{equation}
where $\rho^{in}$ and $\rho^{out}$ are the in-state and out-state's reduced density matrix. The change in entropy is then a function of various coupling constants that enters the amplitude. 

We will construct the tree level eikonal phase in this section, as in figure \ref{fig:Scattering_Graph}, and give
an explanation of why minimal coupling amplitude gives minimal entropy increment in the
purely gravitational case.

\subsection{Constructing the eikonal phase} 
In this subsection, we construct the $2 \rightarrow 2$ scattering amplitude with generic massless particle exchange in the eikonal limit. We set particle $1$ and $1^{\prime}$'s mass to be $m_a$, $2$ and $2^{\prime}$'s mass to be $m_b$. Working in the center of mass frame, where the momenta are parameterized as
\begin{equation}
p_1 = \left(E_a, 	\vec{p}+\frac{\vec{q}}{2} \right), \; 
p_2 = \left(E_b, -	\vec{p}-\frac{\vec{q}}{2} \right), \;
p_{1^{\prime}} = \left(E_a, \vec{p}-\frac{\vec{q}}{2} \right), \; 
p_{2^{\prime}} = \left(E_b, -\vec{p}+\frac{\vec{q}}{2} \right).
\end{equation}
The eikonal limit  then corresponds to the exchange momentum $q^2 \rightarrow 0$. This limit allows us to write the $2 \rightarrow 2$ amplitude with graviton, massless spin 1 and massless scalar exchange as a sum over products of two three point amplitudes:
\begin{equation}\label{eq:Gen_M4}
M_4\big|_{q^2 \rightarrow 0} = \frac{1}{q^2} \Big[
 \underbrace{(M_3^{+2}M_3^{-2}+ M_3^{-2}M_3^{+2})}_{\text{graviton}} + 
 \underbrace{(M_3^{+1}M_3^{-1}+ M_3^{-1}M_3^{+1})}_{\text{massless spin1 }} + 
 \underbrace{(M_3^{\phi}M_3^{\bar{\phi}} + M_3^{\bar{\phi}}M_3^{\phi})}_{\text{massless scalar}}   \Big].
\end{equation}

The three point amplitude with a massless particle of helicity $h$ and two massive particles with spin $s$ in full generality is written as \citep{Arkani-Hamed:2017jhn}
\begin{equation}\label{eq:Gen_M3}
\small
\begin{split}
M_3(\BS{1}^s, \BS{1}^{\prime s}, q^{+|h|})
&=  \frac{\kappa m^2}{2}x^{|h|} R \sum_{i=1}^{2s}g_i \left(\frac{\AB{\BS{1} \BS{1}^{\prime}}}{m}\right)^{2s-i}\left(x \frac{\AB{\BS{1}q}\AB{q\BS{1}^{\prime}}}{m^2}\right)^{i}
= \frac{\kappa m^2}{2}x^{|h|} R \sum_{n=0}^{2s} \frac{C_n}{n!} \left(- \frac{q\cdot \mathbb{S}}{m}\right)^{n} \\
M_3(\BS{1}^s, \BS{1}^{\prime s}, q^{-|h|})
&=  \frac{\kappa m^2}{2}\frac{1}{x^{|h|}} R \sum_{i=1}^{2s}g_i \left(\frac{\SB{\BS{1}\BS{1}^{\prime}}}{m}\right)^{2s-i}\left(\frac{1}{x} \frac{\SB{\BS{1}q}\SB{q\BS{1}^{\prime}}}{m^2}\right)^{i}
= \frac{\kappa m^2}{2} \frac{1}{x^{|h|}} R \sum_{n=0}^{2s} \frac{C_n}{n!} \left( + \frac{q\cdot \mathbb{S}}{m}\right)^{n},
\end{split}
\end{equation}
where $\kappa = \sqrt{32\pi G}$, the $x$ factor is defined by the 3pt kinematics: $m x \RAB{q}  = p_1 \RSB{q}$. Here we discuss the coupling constants $R$ in the three cases we will encounter in this paper:
\begin{itemize}[leftmargin=*]
\item Kerr solution: we only have contributions from graviton ($|h| = 2$) and we set $R=1$.
\item Kerr Newman solution:
We have both gravitational and electromagentic couplings. Note that for photons with $|h| = 1$, the usual coupling constant is $\sqrt{2}Q$ where $Q$ is the electric charge of the particle. However for the Kerr solution, $Q$ will be bounded by the extremal condition $\frac{Q}{m\kappa}\leq1$. Thus near the extremal limit we will have $Q = \kappa m R$, where $R$ is the ratio of charge over the extremal bound.
\item $N = 4$ supergravity: 
For $N = 4$ supergravity with non-zero central charge $Z$, besides graviton and graviphoton, we also have non-zero dilaton profile. When written in terms of superamplitude, the latter two will be related to the gravitational coupling by susy Ward identities, which instill $R$ with dependence on the central charge $Z$.
\end{itemize}
The $C_n$ coefficients in eq.\eqref{eq:Gen_M3} are termed Wilson coefficients, which was originally introduced in the 1-particle EFT\citep{Porto:2008jj, Levi:2015msa}. For gravitational coupling, $C_0 = C_1 = 1$ are universal for all spinning objects. Different spinning gravitating objects can be distinguished by $C_{n>2}$, where it was shown in \citep{Chung:2019duq} that these Wilson coefficients are mapped to the coupling constants $g_i$ in the on-shell basis through the relation:

\begin{equation}\label{gC_map}
g_i = \frac{1}{2^i} \binom{2s}{i}\sum_{n=0}^{i}(-1)^n \binom{i}{n} C_n.
\end{equation}
The spin vector $\mathbb{S}$ is given by
\begin{equation}
\mathbb{S}_{a, b}^{\mu} = \left( \frac{\vec{p}_{a, b} \cdot \vec{\Sigma}}{m}, \vec{\Sigma} + \frac{\vec{p}_{a, b} \cdot \vec{\Sigma}}{ m_{a, b}(m_{a, b} + E_{a, b}) } \vec{p}_{a, b}  \right),
\end{equation}
where $\vec{\Sigma}$ is the spin-$s$ rest frame spin operator satisfying the commutation relation $[\Sigma_i, \Sigma_j] = i \epsilon_{ijk}\Sigma_k$, and we use the convention $\vec{p}_a = \vec{p}_1$, $\vec{p}_b = \vec{p}_2$. 

Given eq.\eqref{eq:Gen_M4}, \eqref{eq:Gen_M3} we define the shorthand notation:
\begin{equation}
W_{i,\eta} \equiv \sum_{n=0}^{2s} \frac{C_{i,n}}{n!} \left(-\eta \frac{q\cdot \mathbb{S}_i}{m_i}\right)^{n}.
\end{equation}
The amplitude can be written as\footnote{We assume the coupling of $\phi$ and $\bar{\phi}$ to be the same.}
\begin{equation}\label{eq:Gen_Eik}
\small
\begin{split}
&\quad
M_4\big|_{q^2 \rightarrow 0}\\
&=
\frac{\kappa^2 m_a^2 m_b^2}{4q^2} 
\left[
\sum_{\eta = \pm 1} \left(\frac{x_a^2}{x_b^2}\right)^{\eta} 
W_{a,\eta}^{(g)}W_{b,\eta}^{(g)}  +
R_a^{(\gamma)} R_b^{(\gamma)} \sum_{\eta = \pm 1} \left(\frac{x_a}{x_b}\right)^{\eta} 
W_{a,\eta}^{(\gamma)}W_{b,\eta}^{(\gamma)} +
R_a^{(\phi)}R_b^{(\phi)}\left(W_{a,1}^{(\phi)}W_{b,1}^{(\bar{\phi})} + W_{a,-1}^{(\bar{\phi})}W_{b,-1}^{(\phi)}\right)
\right]\\
& =
\frac{\kappa^2 m_a^2 m_b^2}{4q^2} 
\left[
\sum_{\eta = \pm 1} e^{2\eta w} 
W_{a,\eta}^{(g)}W_{b,\eta}^{(g)}  -
R_a^{(\gamma)} R_b^{(\gamma)}\sum_{\eta = \pm 1} e^{\eta  w} 
W_{a,\eta}^{(\gamma)}W_{b,\eta}^{(\gamma)} +
R_a^{(\phi)}R_b^{(\phi)}\left( W_{a,1}^{(\phi)}W_{b,1}^{(\bar{\phi})} + W_{a,-1}^{(\bar{\phi})}W_{b,-1}^{(\phi)}\right)
\right],
\end{split}
\end{equation}
where the parantheses in the superscrists denote the exchanged particle and we have used the identity
\begin{equation}
\frac{x_a}{x_b} = -e^w, \quad \cosh w \equiv \frac{p_a \cdot p_b}{m_a m_b}.
\end{equation}

Finally, we dress the amplitude with the Hilbert space matching factors $U_a$ and $U_b$, where $U_a$ and $U_b$ are given by 
\begin{equation}
U_{a,b} = \exp \left[ -i \frac{m_a m_b \mathbb{E}_{a,b}}{(m_{a,b} + E)E} \right], 
\quad \mathbb{T}_{a,b}=\frac{q\cdot \mathbb{S}_{a,b}}{m_{a,b}},
\quad \mathbb{E}_{a,b} = \frac{1}{m_a^2 m_b}\epsilon_{\mu\nu\rho\sigma}q^{\mu}p_{a}^{\nu}p_b^{\rho}\mathbb{S}_{a,b}^{\sigma},
\quad E = E_a + E_b.
\end{equation}
Follow \citep{Aoude:2020mlg}, we the amplitude after Hilber space matching to $\mathcal{O}(\mathbb{S}_a^{2s_a}\mathbb{S}_b^{2s_b})$ for $s_a - s_b$ scattering:
\begin{equation}\label{eq:HM_Amplitude}
\begin{split}
M_4 U_a \otimes U_b &= -\frac{16\pi Gm_a^2 m_b^2}{q^2} 
\times \\
&\left\lbrace \sum_{m=0}^{\floor{s_a}} \sum_{n=0}^{\floor{s_b}} A_{2m,2n} \left( \mathbb{T}_{a}^{2m} \otimes \mathbb{T}_{b}^{2n}\right)
+ \sum_{m=0}^{\ceil{s_a}-1} \sum_{n=0}^{\ceil{s_b}-1}
A_{2m+1, 2n+1} \left(\mathbb{T}_{a}^{2m+1}  \otimes \mathbb{T}_{b}^{2n+1}\right)
\right. \\
&+ \frac{m_a^2 m_b}{E} 
\sum_{m=0}^{\ceil{s_a}-1} 
\sum_{n=0}^{\floor{s_b}} A_{2m+1, 2n}  \left(\text{Sym} \left[\mathbb{E}_a \mathbb{T}_{a}^{2m} \right]\otimes \mathbb{T}_{b}^{2n} \right) \\
&\left.+ \frac{m_a m_b^2}{E}
\sum_{m=0}^{\floor{s_a}}\sum_{n=0}^{\ceil{s_b}-1} A_{2m, 2n+1} \left(\mathbb{T}_{a}^{2m} \otimes \text{Sym} \left[\mathbb{E}_b \mathbb{T}_{b}^{2n} \right]  \right)  \right\rbrace,
\end{split}
\end{equation}
where $A_{i,j}$ encodes all the coupling constants of the amplitude. The eikonal phase $\chi$ is then the Fourier transform of eq.\eqref{eq:HM_Amplitude} to the impact parameter space.

\subsection{The ultra-relativistic limit of Kerr} 
We first review the tree level gravitational Eikonal phase  in \citep{Aoude:2020mlg}, which showed that the change in the von Neumann entropy is close to $0$ when the massive particle is minimal coupled to graviton, i.e. $g_i = \delta_{i,0}$ or $C_n = 1$. The gravitational eikonal phase can be obtained by setting $W^{(\gamma)} = W^{(\phi)} = W^{(\bar{\phi})} = 0$ in eq.\eqref{eq:Gen_Eik}. Incorporating the Hilbert space matching factors, the $A_{i,j}$ coefficients up to spin $1$ for gravitational interaction defined in eq.\eqref{eq:HM_Amplitude} are given by
\begin{align}
A_{0,0} &= c_{2w}\,, \quad A_{1,0} = \frac{i(2E r_a c_{w} - m_b c_{2w})}{m_a^2 m_b r_a} \,, \\
A_{1,1} &= \frac{c_{2 w } s_{w }^2 m_a m_b}{E^2 r_a r_b}+c_{2 w }-\frac{2 m_b c_{w } s_{w }^2}{E  r_a}-\frac{2 m_a c_{w }s_{w }^2}{E r_b}\,, \nonumber \\
A_{2,0} &=\frac{C_{a,2} c_{2 w }}{2}+\frac{m_b^2 c_{2 w } s_{w }^2}{2 E^2  r_a^2}-\frac{2 m_b c_{w } s_{w }^2}{E r_a}\,,\nonumber  \\
\begin{split}
A_{2,1} &= 
i \left( \frac{E C_{a,2} c_{w }}{m_a m_b^2}
-\frac{C_{a,2} c_{2 w }}{2  m_b^2 r_b}
+\frac{c_{2 w }}{4E^2  r_a^2r_b}
-\frac{c_{4 w }}{8 E^2  r_a^2 r_b}
-\frac{c_{w }}{2 E r_a m_b r_b}
+\frac{c_{3 w }}{2 E  r_a m_br_b}
\right.
\\
&\qquad
\left.
-\frac{c_{2 w }}{m_a r_a m_b}
-\frac{1}{8 E^2  r_a^2 r_b}
-\frac{c_{w }}{4 E m_a r_a^2}
+\frac{c_{3 w }}{4 E m_a r_a^2}
\right)\,,
\end{split}
\nonumber\\
\begin{split}
A_{2,2} &= 
\frac{C_{a,2} C_{b,2} c_{2 w }}{4}
-\frac{C_{a,2} c_{w } s_{w }^2 m_a}{E r_b}
-\frac{C_{b,2} c_{w }s_{w }^2 m_b}{E r_a }
+\frac{c_{2 w } s_{w }^4m_a^2 m_b^2}{4 E^4 r_a^2 r_b^2}
-\frac{c_{w } s_{w }^4m_a m_b^2}{E^3 r_a^2r_b}
-\frac{c_{w } s_{w }^4m_a^2 m_b}{E^3 r_a  r_b^2}
\\
&
+\frac{c_{2 w } s_{w }^2m_a m_b}{E^2  r_a  r_b}
+\frac{C_{b,2} c_{2 w }s_{w }^2 m_b^2}{4E^2  r_a^2}
+\frac{C_{a,2} c_{2 w } s_{w }^2 m_a^2 }{4 E^2 r_b^2}\,.
 \nonumber
\end{split}
\end{align}

Let's consider the explicit form of the eikonal matrix for spin $1$. While in general the form is complicated, in the ultra-relativistic limit $\chi $ simplifies. At leading order in the $k \equiv \frac{|\vec{p}|}{m}$ expansion.
\begin{equation}
\begin{split}
&
\left.\chi\right|_{k\gg1} = \\ 
&\small
k^2\left(
\begin{array}{ccccccccc}
 4 L & 0 & \frac{4 \left(C_{b,2}-1\right)}{B^2} & 0 & 0 & 0 & \frac{4 \left(C_{a,2}-1\right)}{B^2} & 0 & -\frac{24 \left(C_{a,2}-1\right) \left(C_{b,2}-1\right)}{B^4}
   \\
 0 & 4 L & 0 & 0 & 0 & 0 & 0 & \frac{4 \left(C_{a,2}-1\right)}{B^2} & 0 \\
 \frac{4 \left(C_{b,2}-1\right)}{B^2} & 0 & 4 L & 0 & 0 & 0 & 0 & 0 & \frac{4 \left(C_{a,2}-1\right)}{B^2} \\
 0 & 0 & 0 & 4 L & 0 & \frac{4 \left(C_{b,2}-1\right)}{B^2} & 0 & 0 & 0 \\
 0 & 0 & 0 & 0 & 4 L & 0 & 0 & 0 & 0 \\
 0 & 0 & 0 & \frac{4 \left(C_{b,2}-1\right)}{B^2} & 0 & 4 L & 0 & 0 & 0 \\
 \frac{4 \left(C_{a,2}-1\right)}{B^2} & 0 & 0 & 0 & 0 & 0 & 4 L & 0 & \frac{4 \left(C_{b,2}-1\right)}{B^2} \\
 0 & \frac{4 \left(C_{a,2}-1\right)}{B^2} & 0 & 0 & 0 & 0 & 0 & 4 L & 0 \\
 -\frac{24 \left(C_{a,2}-1\right) \left(C_{b,2}-1\right)}{B^4} & 0 & \frac{4 \left(C_{a,2}-1\right)}{B^2} & 0 & 0 & 0 & \frac{4 \left(C_{b,2}-1\right)}{B^2} & 0 & 4 L
   \\
\end{array}
\right)\\
& +\mathcal{O}(k),
\end{split}
\end{equation}
where we normalize the eikonal matrix by an overall factor of $Gm^2$ and define $L\equiv \log b$, $B \equiv b m$. We can see that all the off diagonal elements vanish at $C_{n}=1$, giving no spin flip. Generalizing to spin-2 eikonal matrix, the off diagonal elements' dependence on Wilson coefficients are more complicated but all of them still vanish at $C_{n}=1$ at leading order in large $k$ expansion. The explicit form of the spin 2 eikonal matrix is presented in the ancillary Mathematica file.

Thus we conclude that the spin-entanglement suppression observed for minimal coupling can be traced back to the vanishing of \emph{all} off diagonal elements of the eikonal phase vanish in leading order in $k$, which means minimal coupling gives \emph{zero spin flip} to the in-state in the ultra-relativistic limit. 

\section{Kerr Newman}\label{sec: KN}
It was shown in \citep{Chung:2019yfs} that Kerr-Newman black holes minimal couples to both graviton and photon. Here we construct the $\mathcal{O}(G^1)$ eikonal phase of two KN black hole scattering with tunable charge to mass ratio defined as
\begin{equation}\label{eq:ctmr_def}
R_i = \frac{2\sqrt{2}Q_i}{\kappa m_i} = \frac{Q_i}{\sqrt{16\pi G} m_i},
\end{equation}
where $Q$ is the charge carried by the black hole. Note that the charge to mass ratio cannot be arbitrarily large due to the extremal condition:
\begin{equation}
R_i \leq 1.
\end{equation}
This motivates us to examine the behavior of the eikonal phase near the extremal limit. 

Note that when we count the powers of $G$, the eikonal phase up to linear order in $G$ would require the computation of both tree and 1-loop amplitudes, because the $GQ^2$ dependent piece in the Kerr-Newman metric comes from a one-loop amplitude as shown in \citep{Chung:2019yfs}. However, we are near the extremal limit where $Q_i \sim \sqrt{G}m_i$, so the tree level photon exachange amplitude scales as $Q_a Q_b \sim Gm_a m_b$ and the 1-loop amplitude scales as $GQ_a^2 \sim G^2m_a^2$, thus we can drop the 1-loop effect leaving only tree amplitudes in our analysis.

In this section, we first compute the impulse experienced by a scalar particle in Kerr-Newman background from both classical equations of motion and amplitude methods, showing that the Kerr-Newman black hole indeed minimal couples to photon and graviton. Then, we construct the eikonal phase and show that the entropy change is minimized when the extremal condition is satisfied.

\subsection{Matching Impulse from classical EOM and amplitudes}
\begin{figure}[H]
\centering
\includegraphics[scale=0.4]{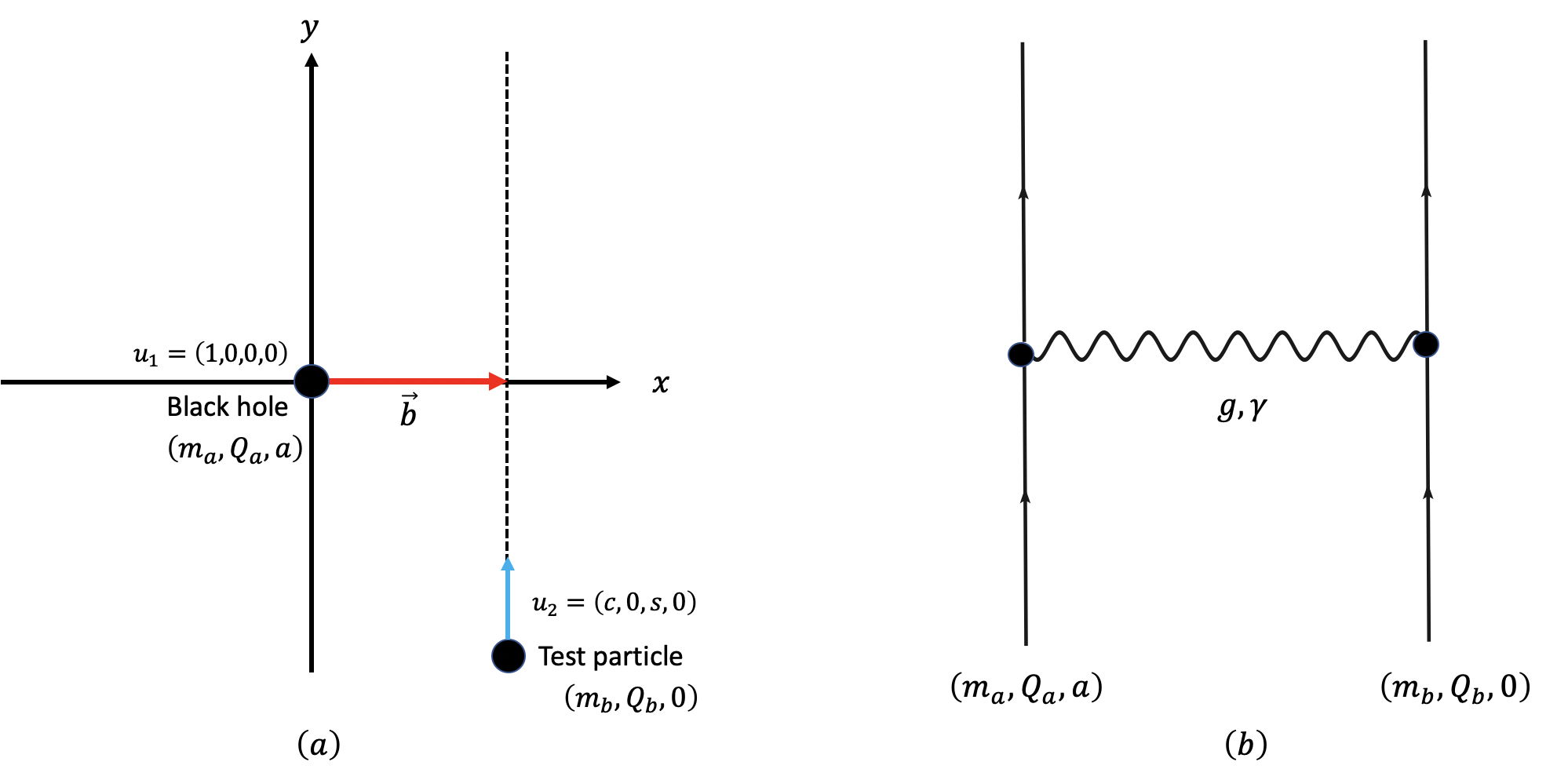}
\caption{(a)The coordinate system set up. We set $z=0$ as the scattering plane and the impact parameter is aligned with the $x$ axis. The black hole sitting at the origin has mass, charge and spin $(m_a, Q_a, a)$ and the scalar test particle has mass, charge $(m_b, Q_b)$. (b)The diagram needed for the amplitude computation. $g$ and $\gamma$ denote graviton amd photon exchange respectively.}
\label{fig:Scattering plane}
\end{figure}

In this subsection, we explicitly compute the impulse experienced by a charged scalar probe in Kerr-Newman background using both the classical EOM and amplitude. On the classical EOM side, we solve for the impulse by integrating the geodesic equation + Lorentz force equation iteratively to linear order in $G$. This result will later be matched to the impusle computed from the amplitude that corresponds to the diagram in figure \ref{fig:Scattering plane}-(b).

\paragraph{Impulse from classical computations}
The impulse of a scalar particle in Kerr Newman background can be computed by the solving the classical EOM, which is the geodesic equation + Lorentz force equation:
\begin{equation}
m_b \frac{d^2 x^{\mu}}{d\tau^2} = -m_b\Gamma^{\mu}_{\nu \rho} \frac{dx^{\nu}}{d\tau}\frac{dx^{\rho}}{d\tau} + Q_b F_a^{\mu\nu} g_{\nu\rho} \frac{dx^{\rho}}{d\tau}.
\end{equation}

As discussed in the beginning of this section, we only take tree level effects into account, so we compute the impulse by iteratively integrating the EOM to leading order in constant velocity approximation:
\begin{equation}
\label{eq:velocity app}
\frac{d x_b^{\mu}}{d \tau} = u_b^{\mu} + \mathcal{O}(G) + \mathcal{O}(Q_b),
\end{equation}
where $u_2^{\mu}$ is the initial proper velocity of the test particle. Then the impulse takes the form
\begin{equation}\label{eq:Impulse_EOM}
\Delta p^{\mu} 
= m_b \int_{-\infty}^{\infty} d\tau \frac{d^2 x^{\mu}}{d\tau^2} 
= m_b \int_{-\infty}^{\infty} d\tau \left( -\Gamma^{\mu}_{\nu \rho}\Big|_{G} u_{b}^{\nu}u_b^{\rho} + Q_b F_{a,\nu}^{\mu}\Big|_{Q_a} u_{b}^{\nu}\right) + \mathcal{O}(G^2) +\mathcal{O}(Q_b^2).
\end{equation}  
For Kerr-Newman background, this iterative solution of the impulse contains tree level effects of interest and also 1-loop effect of order $\mathcal{O}(G Q_a^2)$ coming from the geodesic equation. We compute the impulse from tree effects for the rest of the section and leave the discussion of $\mathcal{O}(G Q_a^2)$ impulse in \citep{Chung:2019yfs}.

We now begin to solve the EOM eq.\eqref{eq:Impulse_EOM}. The set up of the coordinates is shown in figure \ref{fig:Scattering plane}-(a). The scattering plane is the $z=0$ plane and the Kerr-Newman black hole of mass $m_a$ sits at the origin of the scattering plane so its proper velocity is $u_1 = (1,0,0,0)$. For the test particle particle of mass $m_b$, the impact parameter is aligned with the $x$-axis and its proper velocity is: $u_ = (\cosh w, 0, \sinh w, 0) \equiv (c_w, 0, s_w, 0)$, where $w$ is its rapidity. The Christoffel symbol and field strength in eq.\eqref{eq:Impulse_EOM} is computed from the Kerr-Newman metric and gauge field: 
\begin{equation}
\begin{split}
ds^2 &= -\rho^2 \left( \frac{dr^2}{\Delta} + d\theta^2 \right) + \left(dt -a\sin^2\theta d\phi \right)^2 - \frac{\sin^2\theta}{\rho^2}\left[(r^2 + a^2)d\phi - a dt\right]^2,\\
A_{\mu} &=\left(\frac{Q r}{4 \pi  \left(a^2 \cos ^2 \theta+r^2\right)},0,0,-\frac{a Q r \sin ^2(\theta )}{4 \pi \left(a^2 \cos ^2 \theta+r^2\right)}\right),\\
\Delta &= a^2 + \frac{GQ^2}{4\pi}  - 2 G M r + r^2,\\
\rho &= r^2 + a^2 \cos^2 \theta.
\end{split}
\end{equation}
Then the impulse to all orders in spin is computed to be:
\begin{equation}\label{eq:gimpulse}
\begin{split}
\Delta p^x &= -\frac{2 G m_a m_b \left(a s_{2 w}-b c_{2 w}\right)}{\left(a^2-b^2\right) s_w} + \frac{2 G m_a m_b R_a R_b \left(a s_w-b c_w\right)}{\left(a^2-b^2\right) s_w}\\
\Delta p^t &= \Delta p^y = \Delta p^z = 0,
\end{split}
\end{equation}
where we rewrite the charge $Q$ as charge to mass ratio $R$ with the map eq.\eqref{eq:ctmr_def}. 

\subsection{Impulse from amplitude}
The formula for impulse from scattering amplitudes was derived in \citep{Kosower:2018adc}:
\begin{equation}\label{eq:aimpulse}
\Delta p_b^{\mu}=\frac{1}{4m_a m_b}\int \frac{d^4q}{(2\pi)^2}\delta(q\cdot u_1)\delta(q\cdot u_2)e^{-i q\cdot b} i q^{\mu} M_4(1,2\rightarrow 1',2')\mid _{q^2\rightarrow0}.
\end{equation}
In our case of computing the impulse of a scalar test particle in Kerr-Newman background, we take $M_4$ as the four point amplitude of a spin $s$ particle scattering with a scalar particle exchanging a graviton or a photon as shown in figure \ref{fig:Scattering plane}-(b). This amplitude can be simply obtained by setting $W_{b,\eta}^{(g,\gamma)} = 1$ and $R_a^{(\phi)} = R_b^{(\phi)} = 0$ in eq.\eqref{eq:Gen_Eik}, and in the large spin limit, it takes the form:
\begin{equation}
M_4 = \frac{8\pi Gm_a^2 m_b^2}{q^2}\left[ (e^{2w} - R_aR_be^{w})e^{\vec{q}\cdot \vec{a}} +  (e^{-2w} - R_aR_be^{-w})e^{-\vec{q}\cdot \vec{a}} \right].
\end{equation}
Using eq.\eqref{eq:aimpulse}, we follow the description in \citep{Arkani-Hamed:2019ymq}, the impulse along the $x$ axis is
\begin{equation}\label{eq:KN_all_order_imp_from_amp}
\centering
\begin{split}
\Delta p_b^{x} 
&= \frac{2\pi G m_a m_b}{s_w} \int \frac{d^2 \vec{q}}{(2\pi)^2} 
\frac{q^x}{q^2}
\left[
\left(c_{2w} - R_a R_b c_{w}\right) \left(e^{i \vec{q}\cdot \vec{b}_{-}} + e^{i \vec{q}\cdot \vec{b}_{+}}\right) +
\left(s_{2w} - R_a R_b s_{w}\right) \left(e^{i \vec{q}\cdot \vec{b}_{-}} - e^{i \vec{q}\cdot \vec{b}_{+}}\right)
\right]\\
&=
\frac{2\pi G m_a m_b}{s_w} \int \frac{d^2 \vec{q}}{(2\pi)^2} 
\frac{1}{q^2}
\left[
q^x\left(c_{2w} - R_a R_b c_{w}\right) \left(e^{i \vec{q}\cdot \vec{b}_{-}} + e^{i \vec{q}\cdot \vec{b}_{+}}\right) -
iq^z\left(s_{2w} - R_a R_b s_{w}\right) \left(e^{i \vec{q}\cdot \vec{b}_{-}} - e^{i \vec{q}\cdot \vec{b}_{+}}\right)
\right] \\
&= 
-\frac{2Gm_a m_b}{s_w(b^2 - a^2)} 
\left[
\left(c_{2w} - R_a R_b c_{w}\right)b  -
\left(s_{2w} - R_a R_b s_{w}\right)a
\right],
\end{split}
\end{equation}
where we define $\vec{b}_{\pm} = \vec{b} \pm i\vec{a}$. And in the second equality, we  used the identity:
\begin{equation}
q^{\mu} = \frac{i}{s_w}\epsilon^{\mu\nu\rho\sigma}u_{1,\nu}u_{2,\rho} q_{\sigma} \Rightarrow q^x = \frac{i}{s_w}\epsilon^{1023}(-s_w)(-q^z) = -iq^z.
\end{equation}
Comparing eq.\eqref{eq:KN_all_order_imp_from_amp} and eq.\eqref{eq:gimpulse}, we find an exact match to all orders in spin.

\subsection{Eikonal phase}
In this subsection, we observe how the ultra-relativistic behavior of the eikonal phase vary with the charge to mass ratio $R$. We then show that the pattern we observe in the eikonal phase is consistent with the fact that the entropy increment is minimized as the extremal limit is saturated.

Since it was established that the photon and graviton minimal coupling amplitudes capture the spin multipole moments of Kerr-Newman black holes to all orders in spin, the amplitude to arbitrary order in spin can be obtained by setting all the Wilson coefficients to $1$ and $R_a^{(\phi)} = R_b^{(\phi)} = 0$ in eq.\eqref{eq:Gen_Eik}:
\begin{equation}
M_4(1^{s_a}, 1'^{s_a}, 2^{s_b}, 2'^{s_b})\big|_{q^2 \rightarrow 0} = \frac{8\pi G m_a^2 m_b^2}{q^2}
\left[\sum_{\eta = \pm 1} e^{2\eta w} 
W_{a,\eta}^{(g)}W_{b,\eta}^{(g)}  -
R_a R_b\sum_{\eta = \pm 1} e^{\eta  w} 
W_{a,\eta}^{(\gamma)}W_{b,\eta}^{(\gamma)}\right].
\end{equation} 
Incorporating the Hilbert space matching factors, we can compute the eikonal phase to arbitrary orders in spin. 

Here, we first present the spin 1 eikonal phase in the ultra-relativistic limit demonstrating the minimization of spin flip as the extremal bound is saturated. The exact same pattern found in spin 1 persists in higher spin eikonal phase, but we put the explicit form of the spin 2 eikonal matrix in the Mathematica ancillary for brevity. The spin 1 $A_{i,j}$ coefficients defined in eq.\eqref{eq:HM_Amplitude} are given by eq.\eqref{eq:KN_Aij}.

We once again expand the spin 1 eikonal phase in large $k \equiv \frac{|\vec{p}|}{m}$ limit:
\begin{equation}\label{eq:KN_chi}
\begin{split}
&\left.\frac{\chi}{Gm^2}\right|_{k\gg 1} = \chi (k^2) + \chi (k) + \chi (k^0) \\+ \frac{1}{k}
&\tiny
\left(
\begin{array}{ccccccccc}
 0 & -\frac{i \sqrt{2} \left(R^2-2\right)}{B} & -\frac{4}{B^2} & -\frac{i \sqrt{2} \left(R^2-2\right)}{B} & -\frac{8}{B^2} & -\frac{8 i \sqrt{2}}{B^3} & -\frac{4}{B^2}
   & -\frac{8 i \sqrt{2}}{B^3} & 0 \\
 \frac{i \sqrt{2} \left(R^2-2\right)}{B} & 0 & -\frac{i \sqrt{2} \left(R^2-2\right)}{B} & 0 & -\frac{i \sqrt{2} \left(R^2-2\right)}{B} & -\frac{8}{B^2} & 0 &
   -\frac{4}{B^2} & -\frac{8 i \sqrt{2}}{B^3} \\
 -\frac{4}{B^2} & \frac{i \sqrt{2} \left(R^2-2\right)}{B} & 0 & 0 & 0 & -\frac{i \sqrt{2} \left(R^2-2\right)}{B} & 0 & 0 & -\frac{4}{B^2} \\
 \frac{i \sqrt{2} \left(R^2-2\right)}{B} & 0 & 0 & 0 & -\frac{i \sqrt{2} \left(R^2-2\right)}{B} & -\frac{4}{B^2} & -\frac{i \sqrt{2} \left(R^2-2\right)}{B} &
   -\frac{8}{B^2} & -\frac{8 i \sqrt{2}}{B^3} \\
 -\frac{8}{B^2} & \frac{i \sqrt{2} \left(R^2-2\right)}{B} & 0 & \frac{i \sqrt{2} \left(R^2-2\right)}{B} & 0 & -\frac{i \sqrt{2} \left(R^2-2\right)}{B} & 0 & -\frac{i
   \sqrt{2} \left(R^2-2\right)}{B} & -\frac{8}{B^2} \\
 \frac{8 i \sqrt{2}}{B^3} & -\frac{8}{B^2} & \frac{i \sqrt{2} \left(R^2-2\right)}{B} & -\frac{4}{B^2} & \frac{i \sqrt{2} \left(R^2-2\right)}{B} & 0 & 0 & 0 & -\frac{i
   \sqrt{2} \left(R^2-2\right)}{B} \\
 -\frac{4}{B^2} & 0 & 0 & \frac{i \sqrt{2} \left(R^2-2\right)}{B} & 0 & 0 & 0 & -\frac{i \sqrt{2} \left(R^2-2\right)}{B} & -\frac{4}{B^2} \\
 \frac{8 i \sqrt{2}}{B^3} & -\frac{4}{B^2} & 0 & -\frac{8}{B^2} & \frac{i \sqrt{2} \left(R^2-2\right)}{B} & 0 & \frac{i \sqrt{2} \left(R^2-2\right)}{B} & 0 & -\frac{i
   \sqrt{2} \left(R^2-2\right)}{B} \\
 0 & \frac{8 i \sqrt{2}}{B^3} & -\frac{4}{B^2} & \frac{8 i \sqrt{2}}{B^3} & -\frac{8}{B^2} & \frac{i \sqrt{2} \left(R^2-2\right)}{B} & -\frac{4}{B^2} & \frac{i
   \sqrt{2} \left(R^2-2\right)}{B} & 0 \\
\end{array}
\right) \\
&+ \mathcal{O}(k^{-2}),
\end{split}
\end{equation}
where we normalize the eikonal phase by an overall factor of $Gm^2$ and define $B \equiv bm$, $R^2\equiv R_a R_b$ for simplicity. We can see that the $R_a R_b$ dependent off diagonal elements starts at $\mathcal{O}(k^{-1})$. At this order, there are elements that has no dependence on $R_aR_b$ and terms that depends on $R_aR_b$ in the way $(R_a R_b - 2)$. Note that as we are at long distance, the impact parameter $b$ should be larger than the Compton wavelength $\lambda_c \sim m^{-1}$, so that we need to further expand in,
\begin{equation}
B = b m \gg 1\,.
\end{equation}
With this additional condition taken into account, we can see that at $\mathcal{O}(k)^{-1}$, the $R_a R_b - 2$ dependent terms is leading in the $B$ expansion. Since we know that the extremal condition gives an upper bound to the charge to mass ratio: $R_a R_b \leq 1$, the spin flip decreases as the charged BH approaches the extremal condition. 

\begin{figure}[t]
\centering
\includegraphics[scale=0.45]{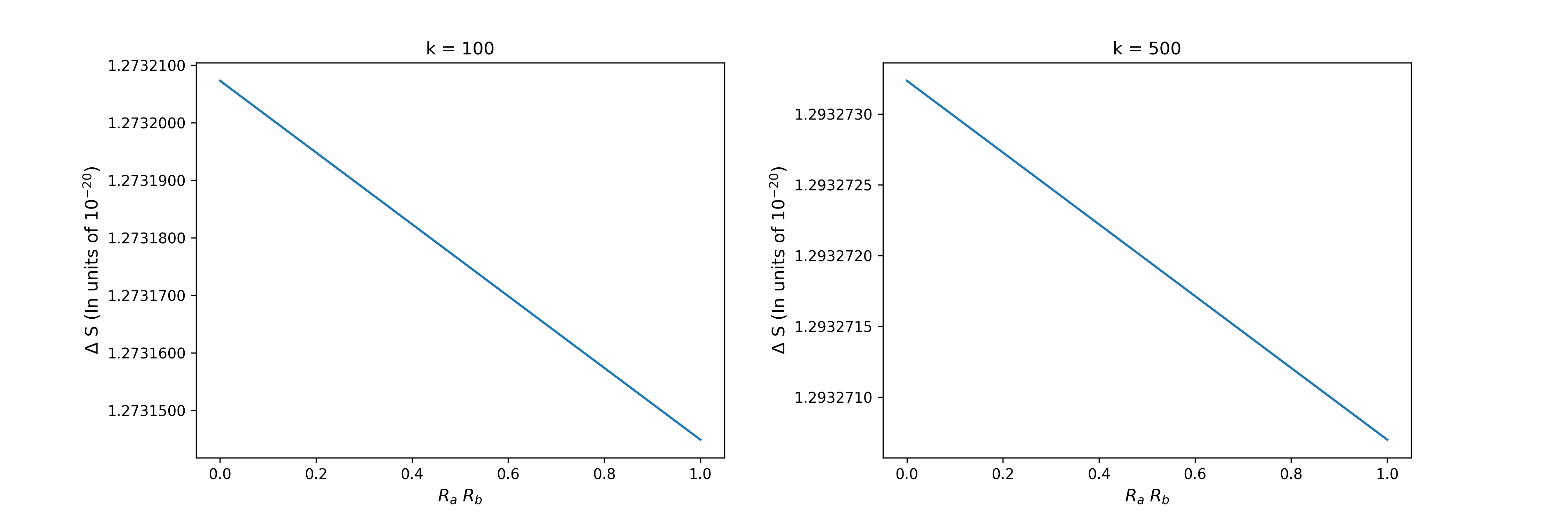}
\caption{The change of von Neumann entropy. We us the kinematics $Gm^2 = 10^{-6}$, $bm = 1000$, $k = \frac{|\vec{p}|}{m} = 100, 500$. The change of von Neumann entropy decreases linearly as the charge to mass ratio increases.}
\label{fig:KN_S_plot}
\end{figure}

We now observe how the change in von-Neumann entropy varies with the product of charge to mass ratios. Here, we use the spin 2 eikonal $S$-matrix to compute the entropy increment. In figure \ref{fig:KN_S_plot}, we plot $\Delta S$ against $R_a R_b$ for various values of large $k$. 
\noindent
From the plots above, we can see that $\Delta S$ decreases as $R_a R_b \rightarrow 1$, which is a reflection of the spin flip minimization pattern we observed in the eikonal matrix eq.\eqref{eq:KN_chi}. Thus we conclude that the spin flip minimizes when extremal limit is saturated.


\section{ $\mathcal{N}=4$ non BPS rotating black holes   }\label{sec: SUSY}

\newcommand{\LRA}[1]{\langle #1  \rangle}
\newcommand{\LRB}[1]{[ #1 ]}

\newcommand{\RA}[1]{| #1 \rangle}
\newcommand{\LA}[1]{\langle #1 |}
\newcommand{\RB}[1]{| #1 ]}
\newcommand{\LB}[1]{[ #1 |}
\newcommand{\etas}[2]{(\eta_{#1}^{#2}\cdot\eta_{#1}^{#2})}
\newcommand{\ttilde}[1]{\tilde{\mathcal{#1}}^{(2)}}
\newcommand{\Math}[1]{\mathcal{ #1}}
\newcommand{\Dotted}[2]{(\mathcal{ #1}\cdot\mathcal{ #2 })}

\newcommand{\ETAsU}[2]{(\eta_1^{#1}\cdot\eta_1^{#2})-(\eta_{1^\prime}^{#1}\cdot\eta_{1^\prime}^{#2})}
\newcommand{\ETAsL}[2]{(\eta_{1 #1}\cdot\eta_{1 #2})-(\eta_{{1^\prime} #1}\cdot\eta_{{1^\prime} #2})}

\newcommand{\tZ}{\theta_Z}
\newcommand{\tZa}{\theta_{Z_{12}}}
\newcommand{\tZb}{\theta_{Z_{34}}}
\newcommand{\CosT}{\sin(\tZa)}
\newcommand{\SinT}{\sin(\tZa)}
\newcommand{\CosTa}{\cos(\tZa)}
\newcommand{\SinTa}{\sin(\tZa)}
\newcommand{\CosTb}{\cos(\tZb)}
\newcommand{\SinTb}{\sin(\tZb)}
\newcommand{\CosTsqr}{\cos^2(\tZa)}
\newcommand{\SinTsqr}{\sin^2(\tZa)}
\newcommand{\SinTwo}{\sin(2\tZ)}

\newcommand{\MmmP}{\Math{M}_{\bold{12}P}}
\newcommand{\MmmPr}{\Math{M}_{res}}
\newcommand{\eqd}{\stackrel{\delta}{=}}

\newcommand{\AAA}{\Math{A}}
\newcommand{\AAAr}{\Math{A}_{res}}
\newcommand{\Ah}{\bold{h}}
\newcommand{\Ag}{\bold{g}}
\newcommand{\Af}{\bold{f}}

\newcommand{\field}[1]{\left|#1\right\rangle}

In this section, we discuss the $\mathcal{N}=4$ supersymmetric rotating black holes. The main difference between supersymmetric black holes and KN black holes is the non-trivial scalar profile. In the first two subsections, we make the connection between the large-spin limit of the component fields in the super multiplet and a non-BPS rotating black hole solution. More precisely, on one hand, we calculate the impulse that a scalar test particle experiences under the influence of a non-BPS black hole solution, which can be derived from the geodesic equation of the test particle. On the other, we take the large-spin limit of a component in the SUSY multiplet, which couples to massless bosonic fields minimally, and calculate the impulse that a scalar test particle experiences through the interactions with this spinning particle. The impulses calculated from these two different approaches match with each other.

After identifying the one-shell amplitude that matches to the 1 PM $\mathcal{N}=4$ blackhole dynamics, we utilize it to construct the eikonal S-matrix of the $2\rightarrow 2$ scattering. Remarkably, we will show that in the ultra-relativistic limit we find that the spin-flipping component vanishes when the BPS  bound is saturated, $Z=2m$.

\subsection{Impulse from geodesic}
\label{subsec:impulse from non BPS}
The bosonic part of the $\mathcal{N}=4$ supergravity action takes the form\citep{Sen:1994eb}: 
\begin{equation}
\label{eq:nonBPSaction}
\small
S = \int d^4 x 
\sqrt{{-}|g|} \left[ 
\frac{{-}R}{16\pi G}{+} 
\frac{1}{2} \partial^{\mu} \phi \partial_{\mu} \phi {+} 
\frac{e^{-2\sqrt{16\pi G} \phi}}{192 \pi}  H^{\mu\nu\rho} H_{\mu\nu\rho} 
- \frac{1}{4} e^{-\sqrt{16\pi G} \phi} F^{\mu\nu}F_{\mu\nu}
\right].
\end{equation}
The non-BPS black hole solution was found in~\citep{Sen:1992ua}, the non-trivial metric, gravi-photon and scalar profile\footnote{Here, we start from the action in \citep{Sen:1992ua} and do the following field redefinitions:  $G_{\mu \nu}\rightarrow e^{\phi}g_{\mu \nu}$, $F_{\mu\nu} \rightarrow \sqrt{4\pi}F_{\mu\nu}$, $Q\rightarrow\frac{Q}{\sqrt{4\pi}}$, and the coordinate shift $\phi \rightarrow \sqrt{16\pi G}\phi$, $r \rightarrow r-\frac{Q^2}{M}$ to obtain eq.\eqref{eq:nonBPSaction}.}:
\begin{align}
\label{eq:SUSYmetric}
\nonumber
A_{\mu}&= \left\{\frac{Q^3-4 \pi  M Q r}{4 \pi  \left(r \left(Q^2-4 \pi  M r\right)-4 \pi  a^2 M
   \cos ^2(\theta )\right)},0,0,\frac{a Q \sin ^2(\theta ) \left(Q^2-4 \pi  M
   r\right)}{4 \pi  \left(4 \pi  a^2 M \cos ^2(\theta )+r \left(4 \pi  M
   r-Q^2\right)\right)}\right\}
\\
\nonumber
g_{\mu\nu}&=\left(
\begin{array}{cccc}
 \alpha (r) & 0 & 0 & \frac{\eta (r)}{2} \\
 0 & -\zeta (r) & 0 & 0 \\
 0 & 0 & -\beta (r) & 0 \\
 \frac{\eta (r)}{2} & 0 & 0 & -\sin ^2(\theta ) \gamma (r) \\
\end{array}\right)\\
\nonumber
\alpha (r)&=\frac{4 \pi  a^2 M \cos ^2(\theta )+(r-2 G M) \left(4 \pi  M
   r-Q^2\right)}{4 \pi  a^2 M \cos ^2(\theta )+r \left(4 \pi  M r-Q^2\right)}\\
\nonumber
\eta (r)&=\frac{4 a G M \sin ^2(\theta ) \left(4 \pi  M r-Q^2\right)}{4 \pi  a^2 M
   \cos ^2(\theta )+r \left(4 \pi  M r-Q^2\right)}\\
\nonumber
\zeta (r)&=\frac{4 \pi  a^2 M \cos ^2(\theta )+r \left(4 \pi  M r-Q^2\right)}{4 \pi  a^2 M-(2 G
   M-r) \left(4 \pi  M r-Q^2\right)}\\
   \nonumber
\beta (r)&=a^2 \cos ^2(\theta )-\frac{Q^2 r}{4 \pi  M}+r^2\\
\nonumber
\gamma (r)&=\frac{2 \pi  a^2 M \cos (2 \theta ) \left(4 \pi  a^2 M-(2 G M-r) \left(4 \pi  M
   r-Q^2\right)\right)}{4 \pi  M \left(4 \pi  a^2 M \cos
   ^2(\theta )+r \left(4 \pi  M r-Q^2\right)\right)}
 \\
 &+\frac{2 \pi  a^2 M (2 G M+3 r) \left(4 \pi  M r-Q^2\right)+8 \pi ^2
   a^4 M^2+r^2 \left(Q^2-4 \pi  M r\right)^2}{4 \pi  M \left(4 \pi  a^2 M \cos
   ^2(\theta )+r \left(4 \pi  M r-Q^2\right)\right)}
\nonumber
\\
e^{\sqrt{16\pi G}\phi}&=\frac{16 \pi ^2 a^2M^2 \cos ^2(\theta )+\left(Q^2-4 \pi  M r\right)^2}{4 \pi  \left(4 \pi 
   a^2M^2 \cos ^2(\theta )+M r \left(4 \pi  M r-Q^2\right)\right)}
\nonumber
\\
 B_{03}&=-\frac{a Q^2 \sin ^2(\theta ) \left(Q^2-4 \pi  M r\right)}{4 \pi  M \left(4 \pi  a^2 M
   \cos ^2(\theta )+r \left(4 \pi  M r-Q^2\right)\right)}.
\end{align}
The extremal limit of this metric is  $\frac{Q}{M}=\sqrt{8\pi G}$. We denote the mass and charge of the black hole as $(m_a, Q_a)$, and the mass and charge of the test particle as $(m_b, Q_b)$ for the following discussion.

Now, we compute the impulse experienced by a charged scalar test particle in non-BPS background using classical EOM. We adopt the approach introduced in \citep{1975ApJ...196L..59E}, where the dependence of the scalar field is incorporated into the mass of the test particle , i.e. $m_b \rightarrow m_b(\phi)$, and the action can be written as the original action plus the world line action of test particle
\begin{align}
\label{eq:SUSYactionwithmatter}
S_{total}&=S+S_{m}\\
\nonumber
S_{m}&=-\int d\tau \left[ m_b(\phi)\sqrt{-g_{\mu \nu}\frac{dx_b^{\mu}}{d\tau}\frac{dx_b^{\nu}}{d\tau}} +Q_b A_{\mu}\frac{dx_b^{\mu}}{d\tau}\right].
\end{align}
In this approach, the EOM for the test particle comes from varying this world line action $S_m$ with respect to $x^{\mu}$:
\begin{equation}
\label{eq:EOM}
\frac{d^2 x^{\mu}}{d\tau^2} = - \Gamma^{\mu}_{\nu\rho} \frac{dx^{\nu}}{d\tau}\frac{dx^{\rho}}{d\tau}+g^{\mu\nu}\frac{\partial_{\nu}m_b(\phi) }{m_b(\phi)}
+\frac{\partial_{\nu}m_b(\phi)}{m_b(\phi)}\frac{dx^{\nu}}{d\tau}\frac{dx^{\mu}}{d\tau}+\frac{Q_b}{m_b(\phi)}F_{\rho \sigma}\frac{dx^{\sigma}}{d\tau}g^{\mu \rho}.
\end{equation}
We compute the impulse as in the section \ref{sec: KN}, expand background to order $\mathcal{O}(G)$ and use constant velocity approximation eq.\eqref{eq:velocity app}.

The exact form $m_b(\phi)$ can be obtained by refering to the method introduced in \citep{Julie:2017rpw}, and the details can be found in appendix \ref{appendix:matching}. In short, by considering the asymptotic behavior of $g_{\mu \nu}$, $A_{\mu}$, $\phi$ above and match with the asymptotic solution of eq.\eqref{eq:SUSYactionwithmatter}, i.e. doing a match between expansion of two solutions at infinity $r\equiv r_\infty +(Gm_a+\frac{Q_a^2}{8\pi  m_a})$, we have the differential equation
\begin{equation}
\label{eq:diffofm}
\frac{1}{\sqrt{16\pi G}}	\frac{d m_b(\phi)}{d\phi}=\frac{Q_b^2}{16\pi  G m_b(\phi)}e^{\sqrt{16\pi G}\phi }.
\end{equation}
This differential equation has the solution:
\begin{equation}
	m_b(\phi)=\sqrt{\text{const.}+\frac{Q_b^2}{8\pi G}e^{\sqrt{16\pi G}\phi}}.
\end{equation}
The constant can be fixed by the boundary condition $m_b(0)=m_b$, which leads to $\text{const.}=m_b^2-\frac{Q_b^2}{8\pi G}$. So, $m_b(\phi)$ takes the form:
\ie
	m_b(\phi)&=\sqrt{m_b^2-\frac{Q_b^2}{8\pi G}+\frac{Q_b^2}{8\pi G}e^{\sqrt{16\pi G}\phi}}
	\\
	&=\sqrt{\frac{Q_b^2 \left(Q_a^4-4 \pi  m_a Q_a^2 r\right)}{32 \pi ^2 G \left(4 \pi  m_a^2a^2 \cos
   (\theta )+m_a r \left(4 \pi  m_a r-Q_a^2\right)\right)}+m_b^2},
\fe
where we substitute $e^{\sqrt{16\pi G}\phi}$ in the last line by \eqref{eq:SUSYmetric}. 

Substituting $g_{\mu \nu}$, $A_{\mu}$, $m_b(\phi)$ to eq.\eqref{eq:EOM} and integrate, we obtain the impulse to all orders in spin\ie
\label{eq:gnonbpsimpulse}
\Delta p_{b}^x&=\frac{2Gm_am_b\left(bc_{2w}-as_{2w}\right)}{(a^2-b^2)s_w}-\frac{Q_aQ_b}{4\pi Gm_am_b}\frac{2Gm_am_b\left(bc_{w}-as_w\right)}{(a^2-b^2)s_w}
\\
&+\frac{Q_a^2Q_b^2}{64\pi^2 G^2m_a^2m_b^2}\frac{2Gm_am_bb}{(a^2-b^2)s_w}.
\fe

\subsection{Impulse from amplitude}

We performed the calculation of the impulse by classical equation of motion in the previous subsection. In this subsection, we want to reproduce the result eq.\eqref{eq:gnonbpsimpulse} from superamplitude, which can be constructed by the on-shell symmetry constraints.

With central charge extension, the supersymmetry generators of a massive particle in $\Math{N}=4$ SUSY have the anti-commutation relations
\ie
\label{eqn: the anti-commutation relations}
&\{ Q^A_{i\alpha}, Q^{\dagger}_{i\dot{\alpha}B} \}=p_{i\alpha\dot{\alpha}}\delta^A_B
\\
&\{ Q^A_{i\alpha}, Q^B_{i\beta} \}={1\over2} Z_i^{AB}\epsilon_{\alpha\beta}
\\
&\{ Q^{\dagger}_{i\dot{\alpha}A}, Q^{\dagger}_{i\dot{\beta}B} \}=-{1\over2} Z_{i,AB}\epsilon_{\dot{\alpha}\dot{\beta}}
\;,\fe
where $A$ and $B$ run from $1$ to $4$, and $Z_{iAB}$ is
\ie
Z_{iAB}=
\begin{bmatrix}
0&-Z_{12}&0&0\\
Z_{12}&0&0&0\\
0&0&0&-Z_{34}\\
0&0&Z_{34}&0
\end{bmatrix}
\;.\fe
With central charge extension, the theory is no longer symmetric under $SU(4)$, but is broken into a $SU(2)\otimes SU(2)$ subgroup, with $\epsilon_{AB}|_{A,B=1,2}$ and $\epsilon_{AB}|_{A,B=3,4}$ being the invariants.

In $\Math{N}=4$ supergravity, the massive scalar multiplet is
\ie
\label{eqn: massive spectrum}
{\Psi^{0}}=&\cdots
+\frac{1}{4}\eta_I^{A^\prime}\eta^{IA^{\prime\prime}}\eta_{JA^\prime}\eta_{A^{\prime\prime}}^J\field{0}
+\frac{1}{2}\eta_I^{A^\prime}\eta_{KA^{\prime}}\eta_{J}^{A^{\prime\prime}}\eta_{A^{\prime\prime}}^K\field{1}^{(IJ)}\\
&+\eta_I^{A^\prime}\eta_{JA^{\prime}}\eta_{K}^{A^{\prime\prime}}\eta_{LA^{\prime\prime}}\field{2}^{(IJKL)}
+\cdots,
\fe
where $A^\prime=1,2$, $A^{\prime\prime}=3,4$. The multiplet contains several R-symmetry singlets, including 1 spin-2 field, 5 spin-1 fields, and 5 spin-0 fields. In addition, there are two massless multiplets. One of them ranges from helicity $+2$ to helicity $0$, and the other ranges from helicity $0$ to helicity $-2$,
\ie
\label{eqn: spectrum}
{\Psi^{+2}}
=&\field{+2}
+\eta^{A^\prime}\field{+\tfrac{3}{2}}_{A^\prime}
+\eta^{A^{\prime\prime}}\field{+\tfrac{3}{2}}_{A^{\prime\prime}}\\
+&\eta^{A^\prime}\eta_{A^{\prime}} \field{+1}_{12}
+\eta^{A^{\prime\prime}}\eta_{A^{\prime\prime}} \field{+1}_{34}
+\eta^{A^\prime}\eta^{A^{\prime\prime}} \field{+1}_{A^\prime A^{\prime\prime}}\\
+&\eta^{A^{\prime\prime}}\eta_{A^{\prime\prime}}\eta^{A^{\prime}}\field{+\tfrac{1}{2}}_{A^{\prime}}
+\eta^{A^\prime}\eta_{A^\prime}\eta^{A^{\prime\prime}}\field{+\tfrac{1}{2}}_{A^{\prime\prime}}
+\eta^{A^{\prime}}\eta_{A^{\prime}}\eta^{A^{\prime\prime}}\eta_{A^{\prime\prime}}\field{0}
\\
{\Psi^{0}}
=&\field{0}
+\eta^{A^\prime}\field{-\tfrac{1}{2}}_{A^\prime}
+\eta^{A^{\prime\prime}}\field{-\tfrac{1}{2}}_{A^{\prime\prime}}\\
+&\eta^{A^\prime}\eta_{A^{\prime}} \field{-1}_{12}
+\eta^{A^{\prime\prime}}\eta_{A^{\prime\prime}} \field{-1}_{34}
+\eta^{A^\prime}\eta^{A^{\prime\prime}} \field{-1}_{A^\prime A^{\prime\prime}}\\
+&\eta^{A^{\prime\prime}}\eta_{A^{\prime\prime}}\eta^{A^{\prime}}\field{-\tfrac{3}{2}}_{A^{\prime}}
+\eta^{A^\prime}\eta_{A^\prime}\eta^{A^{\prime\prime}}\field{-\tfrac{3}{2}}_{A^{\prime\prime}}
+\eta^{A^{\prime}}\eta_{A^{\prime}}\eta^{A^{\prime\prime}}\eta_{A^{\prime\prime}}\field{-2}
\;,\fe
where $A^\prime=1,2$, $A^{\prime\prime}=3,4$.

\tikzset{
  photon/.style={decorate, decoration={snake}, draw=black},
  fermion/.style={draw=black, postaction={decorate},decoration={markings,mark=at position .55 with {\arrow{>}}}},
}

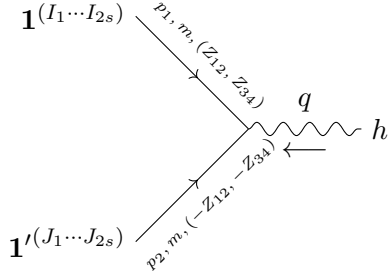
\begin{figure}[H]
\label{fig: 3-pt vertex}
\centering
\begin{tikzpicture}[node distance=0.5cm and 0.5cm]
\coordinate[] (v1);
\coordinate[right=of v1] (v2);
\coordinate[right=of v2] (v3);
\coordinate[right= of v3] (v4);
\coordinate[above left = 1.5cm and 1.5cm of v1,label=left :$\bold{1}^{(I_1\cdots I_{2s})}$ ] (e1);
\coordinate[below left = 1.5cm and 1.5cm of v1,label=left :$\bold{1^\prime}^{(J_1\cdots J_{2s})}$ ] (e2);
\draw[fermion] (e1) -- (v1) node[midway, sloped, above=0.1cm,font=\footnotesize] {\tiny$p_1,m,(Z_{12},Z_{34})$};
\draw[fermion] (e2) -- (v1) node[midway, sloped, below=0.1cm,font=\footnotesize] {\tiny$p_2,m,(-Z_{12},-Z_{34})$};
\draw[photon] (v1) -- (v4) node[midway,below=0.1cm] {$\longleftarrow$} node[right] {$h$  } node[midway, above=0.1cm]{$q$};
\end{tikzpicture}
\caption{A 3-pt superamplitude of two massive spin-s multiplets with equal mass $m$ and a massless particle with helicity $h$ multiplet.}
\end{figure}

Consider a 3pt superamplitude in $\mathcal{N}=4$ supersymmetry, which is shown in Figure \ref{fig: 3-pt vertex}, where two of the external legs are massive (leg $1$ and leg $1^\prime$), having spin-$s$ states being their vacuum states, and the other (leg $q$) is a massless multiplet with helicity $h$. The two massive legs have equal masses $m$ and opposite central charges $Z_{AB}$ and $-Z_{AB}$.

Since the full $SU(4)$ symmetry is broken into $SU(2)\otimes SU(2)$, the superamplitude in $\Math{N}=4$ can be written as a product of two superamplitudes in $\Math{N}=2$. From \citep{Chen:2021hjl}, the SUSY invariants in $\Math{N}=2$ supersymmetry are
\ie
\label{S1: the first solution}
S_{12}^{(1)}(\theta_{Z_{12}})
=&\frac{\CosTsqr}{4}\LRA{\Math{Q}^{A} \Math{Q}_q^{B}}\LRA{\Math{Q}_{A} \Math{Q}_{qB}}
\\
&+\frac{\CosT \SinT}{6x}\LRA{\Math{Q}^{A} \Math{Q}^{B}}\LRB{\tilde{\Math{Q}}_{A} \tilde{\Math{Q}}_{B}} \left( \eta_q\cdot\eta_{q} \right)
\\
&+\frac{\SinTsqr}{12x^2} \LRB{\tilde{\Math{Q}}^{A} \tilde{\Math{Q}}_{q}^{B}} \LRB{\tilde{\Math{Q}}_{A} \tilde{\Math{Q}}_{qB}} \left(\ETAsU{C}{D} \right)\left(\ETAsL{C}{D} \right)
\\
&+\frac{\CosT}{3}\LRA{\Math{Q}^{A} \Math{Q}^{B}}\LRA{\Math{Q}_{A} \Math{Q}_{qB}}
\\
&-\frac{2\SinT}{9x}\LRA{\Math{Q}^{A} \Math{Q}^{B}}  \LRB{\tilde{\Math{Q}}_{A} \tilde{\Math{Q}}_{q}^{C}}\left(\ETAsL{B}{C}\right)
\\
&+\frac{1}{12}\LRA{\Math{Q}^{(A} \Math{Q}^{B)}}\LRA{\Math{Q}_{A} \Math{Q}_{B}}
\;,\fe
\ie
S_{12}^{(2)}(\theta_{Z_{12}})
=&\frac{1}{12x}\LRB{\tilde{\Math{Q}}^{A} \tilde{\Math{Q}}^{B} }\LRB{\tilde{\Math{Q}}_A\tilde{\Math{Q}}_B}\left( \eta_q\cdot\eta_{q} \right)
\\
&-\frac{2\CosT}{9x}\LRB{\tilde{\Math{Q}}^A \tilde{\Math{Q}}^{B}}   \LRB{\tilde{\Math{Q}}_A\tilde{\Math{Q}}_q^{C}} \left( \ETAsL{B}{C} \right)
\\
&+\frac{\SinT}{3} \LRB{\tilde{\Math{Q}}^{A}\tilde{\Math{Q}}^{B}}\LRA{\Math{Q}_A\Math{Q}_{qB}}
\\
&+\frac{\SinTsqr x}{2}\LRA{\Math{Q}^Aq}\LRA{\Math{Q}_Aq}
\\
&+\frac{\CosT\SinT}{6}\LRA{\Math{Q}^{A}\Math{Q}^{B}}\LRB{\tilde{\Math{Q}}_A\tilde{\Math{Q}}_B}
\\
&+\frac{\CosTsqr}{6x} \LRB{\tilde{\Math{Q}}^Aq} \LRB{\tilde{\Math{Q}}_Aq} \left( \ETAsU{B}{C} \right) \left( \ETAsL{B}{C} \right)
\;,\fe
where $A,B,C,D$ run from $1$ to $2$, and
\ie
&\Math{Q}_{\alpha}^A\equiv\lambda_{1\alpha}^I\eta_{1I}^A+\lambda_{1^\prime\alpha}^I\eta_{1^\prime I}^A\;\;;\;\;
\tilde{\Math{Q}}_{\dot{\alpha}}^A\equiv\tilde{\lambda}_{1\dot{\alpha}}^I\eta_{1I}^A-\tilde{\lambda}_{1^\prime\dot{\alpha}}^I\eta_{1^\prime I}^A\\
&\Math{Q}_{q\alpha}^A\equiv\lambda_{q\alpha}\eta_{q}^A\;\;;\;\;
\tilde{\Math{Q}}_{q\dot{\alpha}}^A\equiv\tilde{\lambda}_{q\dot{\alpha}}\eta_{q}^A\\
&\eta_i^A\cdot\eta_i^B\equiv-\frac{1}{2}\epsilon^{IJ}\eta_{iI}^A\eta_{iJ}^B\;\;;\;\;\eta_q\cdot\eta_q\equiv\frac{1}{2}\eta_q^A\eta_{qA}
\;,\fe
and the variable $\theta_{Z_{12}}$ in the equations are related to the central charge $Z_{12}$ by
\ie
\sin(2\theta_{Z_{12}})=\frac{Z_{12}}{2m}
\;.\fe
The subscripts that $S_{12}^{(1)}$ and $S_{12}^{(2)}$ carries indicate which projected $SU(2)$ group they are describing. Note that the degree of the Grassmann variables are not homogeneous in both $S_{12}^{(1)}$ and $S_{12}^{(2)}$. Take $S_{12}^{(1)}$ as an example, the first term $\LRA{\Math{Q}^{A} \Math{Q}_q^{B}}\LRA{\Math{Q}_{A} \Math{Q}_{qB}}$ has $\eta$-degree 4, follows by terms with $\eta$-degree 6,8,4,6,4. $S_{34}^{(1)}$ and $S_{34}^{(2)}$ are defined similarly, with the range of the summed index running from 3 to 4.

The set of superamplitudes that incorporates both the spectrums eq.\eqref{eqn: spectrum} is
\ie
\mathcal{M}(\bold{1}^{2s},\bold{1^\prime}^{2s},{q}^{+2})&= M(\bold{1}^{2s},\bold{1^\prime}^{2s},q^{+2})\cdot   \left( S^{(1)}_{12}(\theta_{Z_{12}}) \cdot S^{(1)}_{34}(\theta_{Z_{34}})\right)\\
\mathcal{M}(\bold{1}^{2s},\bold{1^\prime}^{2s},{q}^{0})&=\bar M(\bold{1}^{2s},\bold{1^\prime}^{2s}q^{0})\cdot \left(S^{(2)}_{12}(\theta_{Z_{12}}) \cdot S^{(2)}_{34}(\theta_{Z_{34}}) \right)
\;,\fe
where $M$ and $\bar M$ are bosonic quantities which carry the spin of the massive multiplets and the helicity of the massless multiplet\footnote{The fact that we can factorize the superamplitude into two parts, one of which is a pure bosonic part that carries the spin and helicity of the external multiplets, and the other is a super invariant part, is proven in \citep{Chen:2021hjl}.}. Recall that in~\citep{Arkani-Hamed:2019ymq}, it was shown that minimally coupled three-point amplitude has the feature that in the classical spin limit, the spin-effects can be mapped to the Janis-Newman shift that relates Kerr solution to Schwarzschild. Since it was shown in \citep{Yazadjiev:1999ce} that the $\mathcal{N}=4$ solution can also be generated from the non-spinning solution via a Janis-Newman shift, this suggests that  $M$ and $\bar M$ are simply:
\ie
&M(\bold{1}^{2s},\bold{1^\prime}^{2s},q^{+2})=\frac{\kappa}{2m^2}x^2 \frac{\LRA{\bold{11^\prime}}^{2s}}{m^{2s}}\\
&\bar M(\bold{1}^{2s},\bold{1^\prime}^{2s},q^{0})=\frac{\kappa}{2m^2} \frac{\LRB{\bold{11^\prime}}^{2s}}{m^{2s}}
\;,\fe 
and the superamplitudes take the form
\ie
\label{eqn: superamplitude -bosonic minimal}
\mathcal{M}(\bold{1}^{2s},\bold{1^\prime}^{2s},{q^{+2}})&=  \frac{\kappa}{2m^{2s+2}} x^2\LRA{\bold{11^\prime}}^{2s}\cdot  \left( S^{(1)}_{12}(\tZ) \cdot S^{(1)}_{34}(\tZ) \right)\\
\mathcal{M}(\bold{1}^{2s},\bold{1^\prime}^{2s},{q^{0}})&= \frac{\kappa}{2m^{2s+2}} \LRB{\bold{11^\prime}}^{2s} \cdot\left( S^{(2)}_{12}(\tZ) \cdot S^{(2)}_{34}(\tZ) \right)
\;.\fe

Since quantum gravity forbids global symmetry, we expect that if we want to describe a non-BPS black hole by a spinning particle, the particle must be a singlet under R-symmetry. We also want all spin effects come from $\LRA{\bold{11^\prime}}^{2s}$, so that they can be mapped to the Janis-Newman shift. Therefore, we have to extract scalar components in massive multiplet eq.\eqref{eqn: massive spectrum}, and get spin-$s$ singlets after combining with $\LRA{\bold{11^\prime}}^{2s}$. It turns out that there are 5 scalar singlets in the massive multiplet, and they all have the same coupling constants when they couple to massless fields. We show how to obtain one of the scalar singlets as an example,

\footnotesize
\ie
\label{eqn:  spin-0 singlet and h=2}
&\frac{1}{4}\prod_{i=1,1^\prime}\frac{\partial}{\partial\eta_{iI}^{A^{\prime}}} \frac{\partial}{\partial\eta_{i}^{IA^{\prime\prime}}} \frac{\partial}{\partial\eta_{iJ{A^{\prime}}}} \frac{\partial}{\partial\eta_{iA^{\prime\prime}}^J}\Bigg|_{\substack{{A^{\prime}}=1,2\\A^{\prime\prime}=3,4}}\left( \frac{\kappa}{2m_1^{2s+2}}  x^2 S^{(1)}_{12} S^{(1)}_{34}\right)
\\
&=\frac{\kappa}{2m_1^{2s-2}} \Bigg\{ x^2
-  x\left( \frac{Z_{12}}{2m_1}\left( \eta_q^{A^{\prime}}\eta_{q{A^{\prime}}} \right) + \frac{Z_{34}}{2m_1}\left( \eta_q^{A^{\prime\prime}}\eta_{q{A^{\prime\prime}}} \right)  \right)
+ \frac{Z_{12}}{2m_1} \frac{Z_{34}}{2m_1} \left( \eta_q^{A^{\prime}}\eta_{q{A^{\prime}}} \right)\left( \eta_q^{A^{\prime\prime}}\eta_{q{A^{\prime\prime}}} \right) \Bigg\}
\;.\fe
\normalsize
From eq.\eqref{eqn:  spin-0 singlet and h=2}, we see that there are 4 fields in massless spin-2 multiplet that the spinning particle can interact with: a graviton ($\field{+2}$ in eq.\eqref{eqn: spectrum}), two gravi-photon fields ($\field{+1}_{12}$ and $\field{+1}_{34}$), and a complex scalar field ($\field{0}$). All of them are singlets under R-symmetry. The values of the central charges fix the relative coefficients of the couplings. In addition, the amplitudes above have their conjugate amplitudes, which can be derived from $\mathcal{M}(\bold{1}^{2s},\bold{1^\prime}^{2s},{q^{0}})$.

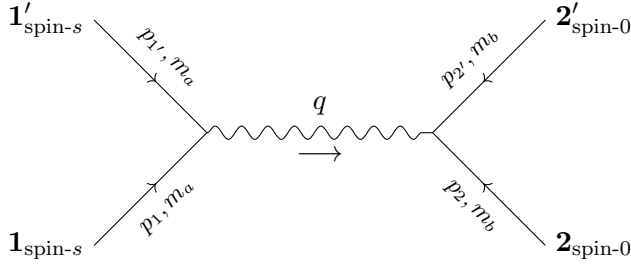
\begin{figure}[H]
\centering
\begin{tikzpicture}[node distance=1cm and 1cm]
\coordinate[] (v1);
\coordinate[right=of v1] (v2);
\coordinate[right=of v2] (v3);
\coordinate[right= of v3] (v4);
\coordinate[above right= 1.5cm and 1.5cm of v4,label=right:$\bold{2}^\prime_{\text{spin-0}}$] (f1);
\coordinate[below right= 1.5cm and 1.5cm of v4,label=right:$\bold{2}_{\text{spin-0}}$] (f2);
\coordinate[above left = 1.5cm and 1.5cm of v1,label=left :$\bold{1}^\prime_{\text{spin-$s$}}$] (e1);
\coordinate[below left = 1.5cm and 1.5cm of v1,label=left :$\bold{1}_{\text{spin-$s$}}$] (e2);
\draw[fermion] (e1) -- (v1) node[midway, sloped, above=0.1cm,font=\footnotesize] {$p_{1^\prime}, m_a$};
\draw[fermion] (e2) -- (v1) node[midway, sloped, below=0.1cm,font=\footnotesize] {$p_1, m_a$};
\draw[photon] (v1) -- (v4) node[midway,below=0.1cm] {$\longrightarrow$} node[midway, above=0.1cm]{$q$};
\draw[fermion] (f1) -- (v4) node[midway, sloped, above=0.1cm,font=\footnotesize] {$p_{2^\prime}, m_b$};
\draw[fermion] (f2) -- (v4) node[midway, sloped, below=0.1cm,font=\footnotesize] {$p_2, m_b$};
\end{tikzpicture}
\caption{A probe scalar particle interacting with a non-BPS spinning particle. For particle $\bold{1}$, the central charges are $(Z_{a12},Z_{a34})$, while for particle $\bold{2}$, the central charges are $(Z_{b12},Z_{b34})$.}
\label{4ptgraph}
\end{figure}

Let's now move on to 4pt superamplitudes, which can be built from 3pt superamplitudes by performing super sum on the Grassmann variables. Consider a scalar test particle (labeled by $2$ and $2^\prime$) with mass $m_b$ and central charge $(Z_{b12},Z_{b34})$ interacting with a spin-$s$ R-charge singlet (labeled by $1$ and $1^\prime$) with mass $m_a$ and central charge $(Z_{a12},Z_{a34})$, as shown in Figure \ref{4ptgraph}. They interact through a series of exchanged massless particles, including graviton, gravi-photon, and scalar. We first calculate the scattering amplitude, and then determine the impulse by eq.\eqref{eq:aimpulse}. We further choose $Z_{a12}=Z_{a34}=Z_a$ and $Z_{b12}=Z_{b34}=Z_b$.

The 4pt amplitude of spin-s particle 1 and a scalar particle 2, illustrated in Figure \ref{4ptgraph}, is
\ie
\label{eqn: BH 4pt amplitude}
M_{4}(\bold{1,2,1',2'})&=\frac{8\pi Gm_a^2m_b^2}{m_a^{2s} q^2}\Big[
\left(\frac{x_{a}}{x_{b}}\right)^2\LRA{\bold{11'}}^{2s}
+\left(\frac{x_{b}}{x_{a}}\right)^2\LRB{\bold{11'}}^{2s}\\
&+\frac{Z_a Z_b}{2m_a m_b}\left(\frac{x_{a}}{x_{b}}\LRA{\bold{11'}}^{2s}
+\frac{x_{b}}{x_{a}}\LRB{\bold{11'}}^{2s}\right)\\
&+\frac{Z_a^2 Z_b^2}{16m_a^2 m_b^2}\left(\LRA{\bold{11'}}^{2s}+\LRB{\bold{11'}}^{2s}\right)\Big]+\Math{O}(q^0)
\;.\fe
We can project this amplitude in the basis eq.\eqref{eq:Gen_M4}, and obtain the amplitude
\begin{equation}\label{eq:SUSYinspinspace}
\begin{split}
M_4
& =
8\pi G\frac{m_a^2m_b^2}{q^2} 
\left[
\sum_{\eta = \pm 1} e^{2\eta w} W_{g}(\eta \mathbb{T})  -
\frac{Z_a Z_b}{2m_a m_b}\sum_{\eta = \pm 1} e^{\eta  w} W_{\gamma}(\eta \mathbb{T}) +\frac{Z_a^2 Z_b^2}{16m_a^2m_b^2}
\sum_{\eta=\pm 1}W_s( \mathbb{\eta T}) 
\right].
\end{split}
\end{equation}

The impulse along the $x$-axis can be calculated from \eqref{eq:aimpulse}. The the spin-2 and spin-1 exchange have been computed in eq.\eqref{eq:KN_all_order_imp_from_amp}, so we are left with the scalar exchanging part,
\begin{equation}\label{eq:SUSY_all_order_imp_from_amp_scalar}
\centering
\begin{split}
\Delta p_{b(\phi)}^{x}
&= \frac{2\pi G m_a m_b}{s_w} \int \frac{d^2 \vec{q}}{(2\pi)^2} 
\frac{q^x}{q^2}
\left(\frac{Z_a^2 Z_b^2}{16m_a^2m_b^2}\right) \left(e^{i \vec{q}\cdot \vec{b}_{-}} + e^{i \vec{q}\cdot \vec{b}_{+}}\right)
= -\frac{2Gm_a m_b}{s_w(b^2 - a^2)} 
\left(\frac{Z_a^2 Z_b^2}{16m_a^2m_b^2}\right)b.
\end{split}
\end{equation}
Including the contribution from graviton and gravi-photon exchanges, the impulse takes the form\footnote{Note that the scalars and gravi-photons are complex pairs, and their relations to the real fields takes the form \ie \phi^{\text{(complex)}}=\frac{\phi_1^{\text{(real)}}+i\phi_2^{\text{(real)}}}{\sqrt{2}}  \;.\fe As a result, the calculations should take the field normalization factor $\frac{1}{\sqrt{2}}$ into account.}
\ie
\label{eq:SUSY_all_order_imp_from_amp}
\Delta p_{2}^x&=\frac{2Gm_am_b\left(bc_{2w}-as_{2w}\right)}{(a^2-b^2)s_w}-\frac{Z_aZ_b}{2m_am_b}\frac{2Gm_am_b\left(bc_{w}-as_w\right)}{(a^2-b^2)s_w}
\\
&+\frac{Z_a^2Z_b^2}{16m_a^2m_b^2}\frac{2Gm_am_bb}{(a^2-b^2)s_w}
\;.\fe

The extremal limit for non-BPS particle is $Z_i=2m_i$, by comparing with the extremum of charge to mass ratio of a charged black hole $Q_i/m_i=\sqrt{8\pi G}$, we can identify $Q_i=\sqrt{2\pi G}Z_i$. Comparing eq.\eqref{eq:SUSY_all_order_imp_from_amp} and eq.\eqref{eq:gnonbpsimpulse}, we find the impulse computed from amplitude is compatible to that from the classical equations of motion.

\subsection{Eikonal phase of non BPS black holes}
\begin{figure}[t]
\centering
\includegraphics[scale=0.4]{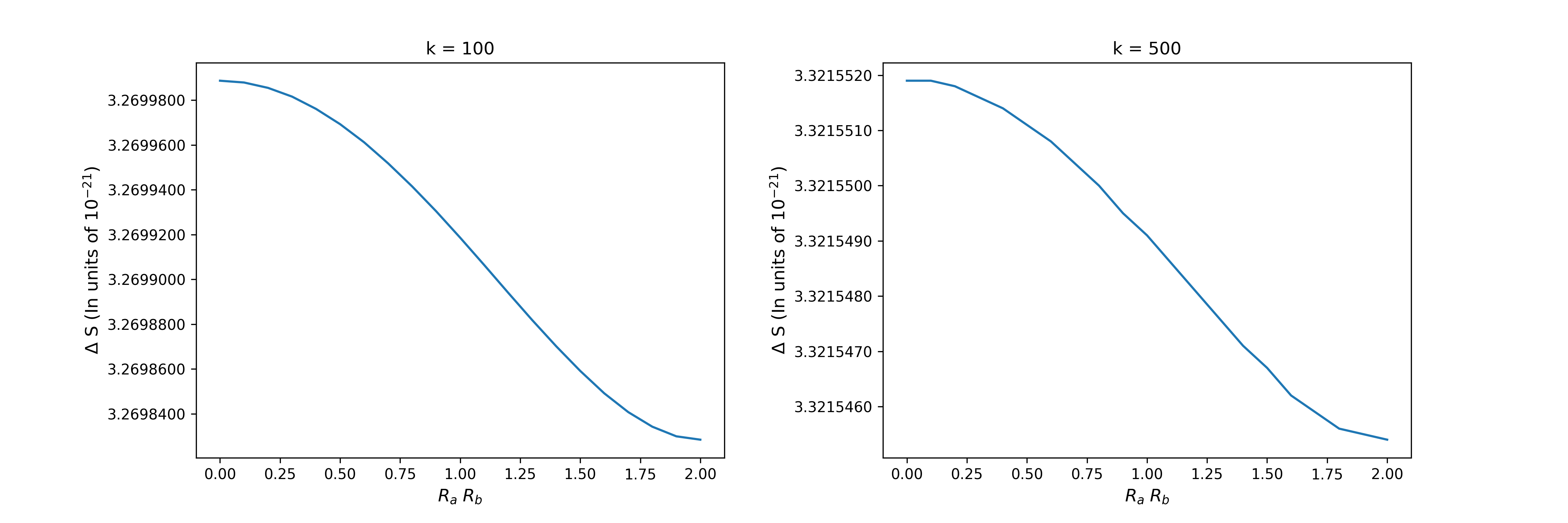}
\caption{The change of von Neumann entropy with same mass $m_a=m_b=m$. We use the kinematics $Gm^2 = 10^{-6}$, $bm=1000$, $k = \frac{|\vec{p}|}{m} = 100, 500$.}\label{fig:BPS_Delta_S}
\end{figure}
\noindent
As shown in\citep{Aoude:2020mlg}, Wilson coefficients $C_2$ has the dominant effect in generating entanglement, hence, we choose spin-1 eikonal phase as an example, the eikonal phase $\chi$ can be written as 
\begin{equation}
\chi=\chi _g+\frac{Z_aZ_b}{2m_am_b}\chi _p+\frac{Z_a^2Z_b^2}{16m_a^2m_b^2}\chi _s,
\end{equation}
where $\chi_i (i=g,p,s)$ denote the exchange particle is graviton, graviphoton and massless scalar respectively. If we expand the eikonal phase $\chi$ with respect to $k=\frac{\abs{\vec{p}}}{m}$,  the matrix elements are
\begin{equation}\label{eq:SUSY_chi}
\small
\begin{split}
&\left.\frac{\chi}{Gm^2}\right|_{k\gg 1} = \chi (k^2) + \chi (k) + \chi (k^0) \\+ \frac{1}{k}
&\small
\left(
\begin{array}{ccccccccc}
 0 & -\frac{i \left(Z^2-4\right)}{\sqrt{2} B} & -\frac{4}{B^2} & -\frac{i
   \left(Z^2-4\right)}{\sqrt{2} B} & -\frac{8}{B^2} & -\frac{8 i \sqrt{2}}{B^3} & -\frac{4}{B^2} &
   -\frac{8 i \sqrt{2}}{B^3} & 0 \\
 \frac{i \left(Z^2-4\right)}{\sqrt{2} B} & 0 & -\frac{i \left(Z^2-4\right)}{\sqrt{2} B} & 0 &
   -\frac{i \left(Z^2-4\right)}{\sqrt{2} B} & -\frac{8}{B^2} & 0 & -\frac{4}{B^2} & -\frac{8 i
   \sqrt{2}}{B^3} \\
 -\frac{4}{B^2} & \frac{i \left(Z^2-4\right)}{\sqrt{2} B} & 0 & 0 & 0 & -\frac{i
   \left(Z^2-4\right)}{\sqrt{2} B} & 0 & 0 & -\frac{4}{B^2} \\
 \frac{i \left(Z^2-4\right)}{\sqrt{2} B} & 0 & 0 & 0 & -\frac{i \left(Z^2-4\right)}{\sqrt{2} B} &
   -\frac{4}{B^2} & -\frac{i \left(Z^2-4\right)}{\sqrt{2} B} & -\frac{8}{B^2} & -\frac{8 i
   \sqrt{2}}{B^3} \\
 -\frac{8}{B^2} & \frac{i \left(Z^2-4\right)}{\sqrt{2} B} & 0 & \frac{i \left(Z^2-4\right)}{\sqrt{2}
   B} & 0 & -\frac{i \left(Z^2-4\right)}{\sqrt{2} B} & 0 & -\frac{i \left(Z^2-4\right)}{\sqrt{2} B} &
   -\frac{8}{B^2} \\
 \frac{8 i \sqrt{2}}{B^3} & -\frac{8}{B^2} & \frac{i \left(Z^2-4\right)}{\sqrt{2} B} & -\frac{4}{B^2}
   & \frac{i \left(Z^2-4\right)}{\sqrt{2} B} & 0 & 0 & 0 & -\frac{i \left(Z^2-4\right)}{\sqrt{2} B}
   \\
 -\frac{4}{B^2} & 0 & 0 & \frac{i \left(Z^2-4\right)}{\sqrt{2} B} & 0 & 0 & 0 & -\frac{i
   \left(Z^2-4\right)}{\sqrt{2} B} & -\frac{4}{B^2} \\
 \frac{8 i \sqrt{2}}{B^3} & -\frac{4}{B^2} & 0 & -\frac{8}{B^2} & \frac{i
   \left(Z^2-4\right)}{\sqrt{2} B} & 0 & \frac{i \left(Z^2-4\right)}{\sqrt{2} B} & 0 & -\frac{i
   \left(Z^2-4\right)}{\sqrt{2} B} \\
 0 & \frac{8 i \sqrt{2}}{B^3} & -\frac{4}{B^2} & \frac{8 i \sqrt{2}}{B^3} & -\frac{8}{B^2} & \frac{i
   \left(Z^2-4\right)}{\sqrt{2} B} & -\frac{4}{B^2} & \frac{i \left(Z^2-4\right)}{\sqrt{2} B} & 0 \\
\end{array}
\right),
\end{split}
\end{equation}
where $Z^2=\frac{Z_a}{m}\frac{Z_b}{m}$. We can see that the coefficients which contain central charges vanish at the extremal limit $Z=2m$, which shows BPS black holes have the maximal spin suppression. 

To be more explicit, we plot the change of von Neumann entropy against the ratio of central charge and mass for spin-1 case to show this property. From figure \ref{fig:BPS_Delta_S}, we can see that for $k \gg 1$, $\Delta S$ is decreasing as the central charge approach to  BPS limit $Z_i=2m_i$, which consistent with our analysis.
\noindent
\section{Conclusion and outlook}
In this paper, we've traced the origin of spin-entanglement suppression for minimal coupling, which reproduces leading PM rotating Kerr dynamics, for minimal coupling observed in~\citep{Aoude:2020mlg}, to the suppression of spin-flipping components in the Eikonal S-matrix. We extend the analysis to couplings that reproduces black hole dynamics with tunable parameters, including Kerr-Newman black hole and $\mathcal{N}=4$ non-BPS black holes. Remarkably, we find that in both cases, the extremal limit corresponds to maximal spin-suppression. This echoes what was observed in strong coupling systems where spin-entanglement suppression occurs at symmetry enhancement points~\citep{Beane:2018oxh}. 

In all the above cases, the minimal coupling used are intimately tied to the Janis-Newman shift, which is known to be a property of the exact solution. Thus it will be likely that the observation found here should persist beyond leading order in PM.  Another motivation that spin-flip contributions are suppressed beyond leading order is the fact that similar effects leads to causality violation for the 2 PM time delay~\citep{AccettulliHuber:2020oou}. In particular the $R^3$ and $FFR$ operators, whose presence leads to causality violation was first pointed out in~\citep{Camanho:2014apa}, leads to potential negative time delay at 2PM which can be traced to precisely the new spin flipping components in the Eikonal phase. It will be interesting to verify this with explicit computations.

\section*{Acknowledgements}
We would especially like to thank Jung-Wook Kim, for discussions on the computation of Eikonal phase for spin effects.  Bt C, Mz C, Yt H, and Mk T is supported by MoST Grant No. 109-2112-M-002 -020 -MY33. Yt H is also supported by Golden
Jade fellowship. 

\newpage
\appendix
\section{$A_{i,j}$ coefficients for the spin-1 Kerr-Newman eikonal phase}
\begin{align}
\small
\begin{split}
A_{0,0} &= c_{2 w }-R_a R_b c_{w }, \;
A_{1,0}  = \frac{i \left(\frac{2 E c_{w }}{m_b}-\frac{c_{2 w }}{r_a}\right)}{m_a^2}+\frac{i R_a R_b \left(m_b c_{w }-E r_a\right)}{m_a^2 r_a m_b}
\end{split}  \nonumber \\ 
\small
\begin{split}
A_{1,1} &= \frac{R_a R_b \left(m_a s_{w }^2 \left(E r_a-m_b c_{w }\right)+E r_b \left(m_b s_{w }^2-E r_a c_{w
   }\right)\right)}{E^2 m_a r_a m_b r_b}\\
&+\frac{m_a s_{w }^2 \left(m_b c_{2 w }-2 E r_a c_{w }\right)+E r_b \left(E r_a
   c_{2 w }-2 m_b c_{w } s_{w }^2\right)}{E^2 m_a r_a m_b r_b}
\end{split}\\
\small
A_{2,0} & = \frac{-4 E r_a m_b c_{w } s_{w}^2+E^2 r_a^2 c_{2 w }+m_b^2 c_{2w} s_{w }^2}{2 E^2 m_a^2 r_a^2}-\frac{R_a R_b
   \left(-2 E r_a m_b s_{w }^2+E^2 r_a^2 c_{w }+m_b^2 c_{w} s_{w}^2\right)}{2 E^2 m_a^2 r_a^2}  \nonumber \\
\small
\begin{split}
A_{2,1}  &=
\frac{i R_a R_b \left(m_a \left(-2 E r_a m_b s_{w }^2+E^2 r_a^2 c_{w }+m_b^2 c_{w } s_{w }^2\right)-E r_b \left(-2
E r_a m_b c_{w }+E^2 r_a^2+m_b^2 s_{w }^2\right)\right)}{2 E^2 m_a^3 r_a^2 m_b^2 r_b} \\
&-\frac{i \left(m_a \left(-4 E r_a m_b
   c_{w } s_{w }^2+E^2 r_a^2 c_{2 w }+m_b^2 c_{2 w} s_{w }^2\right)-2 E r_b \left(-E r_a m_b c_{2 w
   }+E^2 r_a^2 c_{w}+m_b^2 c_{w} s_{w}^2\right)\right)}{2 E^2 m_a^3 r_a^2 m_b^2 r_b} 
\end{split} \nonumber\\
\small
\begin{split}
A_{2,2} &= 
\frac{c_{2 w} s_w^4}{4 E^4 r_a^2 r_b^2}
-\frac{c_w s_w^4}{E^3 m_a r_a^2 r_b}
-\frac{c_w s_w^4}{E^3 r_a m_b r_b^2}
+\frac{c_{2 w}s_w^2}{E^2 m_a r_a m_b r_b}
-\frac{c_w s_w^2}{E m_a^2 r_a m_b}\\
&\quad
-\frac{c_w s_w^2}{E m_a m_b^2 r_b}
+\frac{c_{2 w}}{4 m_a^2 m_b^2}
+\frac{c_{2 w} s_w^2}{4 E^2 m_a^2 r_a^2}
+\frac{c_{2 w} s_w^2}{4 E^2 m_b^2 r_b^2}\\
&\quad
+R_a R_b \left(
-\frac{c_w s_w^4}{4 E^4 r_a^2 r_b^2}
+\frac{c_{2 w} s_w^2}{4 E^3 m_a r_a^2 r_b}
+\frac{c_{2 w}s_w^2}{4 E^3 r_a m_b r_b^2}
-\frac{c_w s_w^2}{E^2 m_a r_a m_b r_b}\right. \\
&\quad
\qquad\qquad
\left.
-\frac{c_w}{4 m_a^2 m_b^2}
-\frac{s_w^2}{4 E^3 m_a r_a^2 r_b}
-\frac{s_w^2}{4E^3 r_a m_b r_b^2}
+\frac{s_w^2}{2 E m_a^2 r_a m_b}\right. \\
&\quad
\qquad\qquad
\left.
+\frac{s_w^2}{2 E m_a m_b^2 r_b}
-\frac{c_w s_w^2}{4 E^2 m_a^2 r_a^2}
-\frac{c_ws_w^2}{4 E^2 m_b^2 r_b^2}
\right)\nonumber
\end{split}
\end{align}
\section{Matching Condition}
\label{appendix:matching}
In this section, we derive the equation \eqref{eq:diffofm} by using the method mentioned in the \ref{subsec:impulse from non BPS}, we expand the solution of two actions $S$, $S_{total}$ which defined in the \ref{subsec:impulse from non BPS} at $r\equiv r_\infty +(GM+\frac{Q^2}{8\pi  M})$, the expansion of the eq.\eqref{eq:SUSYmetric} at $r\equiv r_\infty +(GM+\frac{Q^2}{8\pi  M})$ is 

\begin{equation}
\begin{split}
    g_{tt}&=1-\frac{2Gm_a}{r_ \infty}+\mathcal{O}(r_\infty^{-2}), g_{rr}=1+\frac{2Gm_a}{r_ \infty}+\mathcal{O}(\r_\infty^{-2})\\
    g_{\theta \theta}&=1+\frac{2Gm_a}{r_ \infty}+\mathcal{O}(r_\infty^{-2}), 
    g_{\varphi \varphi}=1+\frac{2Gm_a}{r_ \infty}+\mathcal{O}(r_\infty^{-2})\\
    g_{t \varphi}&=g_{\varphi t}=1+\frac{2Gm_a}{r_ \infty}+\mathcal{O}(r_\infty^{-2})\\
        A_{t}&=\frac{Q_a}{4\pi r_{\infty}}+\mathcal{O}(r_\infty^{-2}), A_{\varphi}=-\frac{a_a Q_a \sin(\theta)^2}{4\pi m_a^2 r}+\mathcal{O}(r_\infty^{-2})\\
	\sqrt{16\pi G}\phi &=-\frac{Q_a^2e^{\sqrt{16\pi G}\phi _{\infty}}}{16\pi m_ar_\infty} +\mathcal{O}(r_\infty^{-2}),\\
\end{split}
\end{equation}
The expansion of the solutions of eq.\eqref{eq:SUSYactionwithmatter} at $r\equiv r_\infty +(Gm_a+\frac{Q_a^2}{8\pi  m_a})$ is 

\begin{equation}\label{eq:KN_Aij}
\begin{split}
    g_{tt}&=1-\frac{2Gm_b(\phi _\infty)}{r_ \infty}+\mathcal{O}(r_\infty^{-2}), g_{rr}=1+\frac{2Gm_b(\phi _\infty)}{r_ \infty}+\mathcal{O}(r_\infty^{-2})\\
    g_{\theta \theta}&=1+\frac{2Gm_b(\phi _\infty)}{r_ \infty}+\mathcal{O}(r_\infty^{-2}), 
    g_{\varphi \varphi}=1+\frac{2Gm_b(\phi _\infty)}{r_ \infty}+\mathcal{O}(r_\infty^{-2})\\
    g_{t \varphi}&=g_{\varphi t}=1+\frac{2Gm_b(\phi _\infty)}{r_ \infty}+\mathcal{O}(r_\infty^{-2})\\
        A_{t}&=\frac{Q_b}{4\pi r_{\infty}}+\mathcal{O}(r_\infty^{-2}), A_{\varphi}=-\frac{a_b Q_b \sin(\theta)^2}{4\pi m_b^2(\phi _\infty) r}+\mathcal{O}(r_\infty^{-2})\\
	\sqrt{16\pi G}\phi &=-\frac{1}{\sqrt{16\pi G} r_\infty}\frac{G dm_b}{d\phi}(\phi _\infty) +\mathcal{O}(r_\infty^{-2}).\\
\end{split}
\end{equation}
By comparing these equations, we have
\begin{align}
m_b(\phi _ \infty)&=m_a
\\
\nonumber
Q_b&=Q_a
\\
\nonumber
	\frac{1}{\sqrt{16\pi G}}\frac{d m_b(\phi)}{d\phi}(\phi _ \infty)&=\frac{Q_a^2}{16\pi G m_a}e^{\sqrt{16\pi G}\phi _\infty}.
\end{align}
Therefore, $Q_b$ is identified the charge $Q$ of black hole, and $m_b(\phi _ \infty)$ correspond to the ADM mass $m_a$, substitute first two equations into third equation, we have equation \eqref{eq:diffofm}. \newpage
\bibliography{mybib}{}
\bibliographystyle{JHEP}

\end{document}